\documentclass[fleqn,usenatbib]{mnras}
\usepackage{amsmath}	
\usepackage{amssymb}	

\usepackage{graphicx}
\usepackage{threeparttable}
\usepackage{array}
\usepackage{multirow}
\usepackage{color}

\title{A Mid-infrared Study of Superluminous Supernovae}

\author[L. Sun et al.]{
Luming Sun$^{1}$\thanks{E-mail: sunluming@ahnu.edu.cn}
, Lin Xiao$^{2,3}$, and Ge Li$^{4}$
\\
$^{1}$Department of Physics, Anhui Normal University, Jiuhuanan Road 189, Wuhu 241002, China\\
$^{2}$Department of Physics, College of Physical Sciences and Technology, Hebei University, Wusidong Road 180, Baoding 071002, China\\
$^{3}$Key Laboratory of High-precision Computation and Application of Quantum Field Theory of Hebei Province, Hebei University,\\ Wusidong Road 180, 071002, Baoding, China\\
$^{4}$Department of Modern Physics, University of Science and Technology of China, Hefei 230026, China
}

\date{Accepted XXX. Received YYY; in original form ZZZ}

\pubyear{}

\begin{document}
\label{firstpage}
\pagerange{\pageref{firstpage}--\pageref{lastpage}}
\maketitle

\begin{abstract}
  We present the mid-infrared (MIR) light curves (LC) of 10 superluminous supernovae (SLSNe) at $z<0.12$ based on WISE data at 3.4 and 4.6 $\mu$m.
  Three of them, including PS15br, SN 2017ens, and SN 2017err show rebrightening which started at 200--400 days and ended at 600--1000 days, indicating the presence of dust.
  In four of the left seven SLSNe, dust emission was detected with monochromatic luminosities of $10^7\sim10^8\ L_\odot$ at epochs of 100--500 days based on MIR colors $W1-W2\sim1$.
  Among the three SLSNe which show rebrightening, we further analysed PS15br and SN 2017ens.
  We modeled the SEDs at 500--700 days, which gives dust temperatures of 600--1100 K, dust masses of $\gtrsim 10^{-2}\ M_\odot$, and luminosities of $10^8\sim10^9$ $L_\odot$.
  Considering the time delay and the huge amount of energy released, the emitting dust can hardly be pre-existing dust heated whether collisionally by shocks or radiatively by peak SLSN luminosity or shock emission.
  Instead, it can be newly formed dust additionally heated by the interaction of circum-stellar medium, indicated by features in their spectra and slowly declining bolometric LCs.
  The dust masses appear to be ten times greater than those formed in normal core-collapse supernovae at similar epochs.
  Combining with the analysis of SN 2018bsz by Chen et al. (2022), we suggest that SLSNe have higher dust formation efficiency, although future observations are required to reach a final conclusion.
\end{abstract}


\section{Introduction}
\label{sec:introduction}

Superluminous supernovae \citep[SLSNe][]{Gal-Yam2012,Gal-Yam2019} are a group of supernovae (SNe) with peak luminosities tens or hundreds higher than those of normal SNe.
They are so rare that only $\sim100$ cases have been discovered.
SLSNe usually take longer to rise and decay than normal SNe: they take one to two months to reach their maximum light after explosion, and take about half a year or longer to evolve into a nebular phase or disappear.
SLSNe can be classified into two types according to their near-peak optical spectra: Superluminous Supernovae I (SLSN-I) with no hydrogen lines, and Superluminous Supernovae II (SLSN-II) with hydrogen lines.
SLSN-I show a blue continuum with weak oxygen and carbon absorption features near the peak time, and show spectra resemble those of SN-Ic in the late time.
SLSN-II show hydrogen and helium broad emission lines, and some of them also show narrow emission lines and hence can be called SLSN-IIn.
The energy sources of SLSNe have not been well understood \citep[see][for a review]{Sukhbold2016}.
The models of the energy sources include pair-instability supernovae \citep[e.g.,][]{Gal-Yam2009}, newly born magnetars \citep[e.g.,][]{Quimby2011}, circum-stellar medium (CSM) interactions \citep[e.g.,][]{Smith2007}, and some others.
Although the mechanism of SLSNe is still uncertain, SLSNe are generally associated with the explosions of massive stars, or even the most massive stars.
In addition, nearby SLSNe often reside in galaxies with low mass and low metallicity \citep[e.g.,][]{Perley2016}.
The SLSN progenitors and hosts are reminiscent of the first generation of stars and galaxies in the early universe \citep{Gal-Yam2009}.
And hence SLSNe might be more common at high redshift than they are now.

Some SLSNe emit in the infrared (IR) band.
From most of SLSNe, the IR emission drops along with their optical emission and can be interpreted as emission from the cooling SLSN photosphere.
While a few SLSNe, which we will enumerate later, show IR excess relative to the expected level of photospheric emission.
SN 2006gy, the first well-studied case of SLSN-IIn, showed a near-infrared (NIR) excess possibly starting around $+$130d (after explosion), rising before $+$400d, dropping after $+$700d, and even being detected around $+$3000d \citep{Smith2008b,Miller2010,Fox2015}.
Another SLSN-II with no narrow lines, SN 2008es, also showed NIR excess on $+$254d and $+$301d (after explosion) \citep{Bhirombhakdi2019}.
The closest SLSN-I to date, SN 2018bsz, showed a NIR excess from $+$232d (after peak) to $+$265d \citep{Chen2022}.
A follow-up mid-infrared (MIR) monitoring revealed an MIR excess which begins from $+$265d at the latest and can still be detected up to $+$535d.

These IR excesses were generally interpreted as emissions from dust around the SLSNe.
However, these works have reached different conclusions about the origin of dust.
\cite{Miller2010} interpreted the NIR excess in SN 2006gy between $+$400d and $+$700d as a light echo from a massive ($\gtrsim10$ $M_\odot$) dusty shell heated by the SN peak luminosity.
\cite{Fox2015} argued that the NIR excess in SN 2006gy at about $+$3000d is not an echo, but an emission from dust that is radiatively heated by late-time CSM interaction.
\cite{Bhirombhakdi2019} interpreted the NIR excess in SN 2008es as emission from newly formed dust.
An important basis for their conclusion is that the H$\alpha$ emission line changes from a symmetric profile in the early time ($+89$d) to a blue-skewed profile in the late time ($+$288d).
The variation in the H$\alpha$ profile is consistent with a scenario that the condensing dust preferentially obscures the emission in the far side \citep[e.g.,][]{Smith2009}.
\cite{Chen2022} demonstrated that the IR excess in SN 2018bsz is difficult to explain by an echo by analysing multi-band monitoring data.
They found that the IR excess can otherwise be interpreted as emission from newly formed dust, and estimated the dust mass to be $10^{-4}$ to $10^{-2}$ $M_\odot$ between $+$265d and $+$535d.
In brief, the dust origin in SLSNe has not been definitively concluded and requires further investigation.

Similarly, the dust origin in most normal SNe remains unclear as the emitting dust can be either pre-existing or newly formed \citep[e.g.,][]{Fox2011,Szalai2013,Tinyanont2016,Szalai2019,Szalai2021}.
This ambiguity hinders further research on some important topics, such as the mass loss of the SN progenitor for the case of pre-existing dust and the dust formation efficiency for the case of newly formed dust.
The same goes for SLSNe.
MIR observations are usually more useful for constraining the dust properties than NIR observations \citep[e.g.,][]{Fox2011}.
Thus we aim for a comprehensive study of the MIR properties of SLSNe, which are still poorly understood.


The Spitzer Space Telescope and the Wide field Infrared Survey Explorer (WISE) telescope \citep{Wright2010} are powerful tools for MIR astronomy.
Most MIR studies of SNe were based on Spitzer data, because of the great spatial resolution ($1.6\arcsec$ in I1 and I2 bands) and the high sensitivity ($5\sigma$ detection limit $I1=19.9$ and $I2=18.7$ with 100-seconds exposure) of the instrument.
While a few MIR studies have used WISE data \citep[e.g.,][]{Prieto2012,Fox2013,Kokubo2019,Tartaglia2020,Moriya2020}.
The WISE telescope has been conducting repetitive all-sky surveys since 2010.
The surveys with the telescope include the initial WISE survey, the NEOWISE survey in 2010 and 2011 \citep{Mainzer2011}, and the NEOWISE-R survey restarted in 2014 after a hiatus between 2011 and 2013 \citep{Mainzer2014}.
The strategy of the surveys results in a regular cadence: a typical target is visited by the telescope every half a year (except for the hiatus), and during each visit 8--25 exposures were taken.
Initially, 4 filters were used, and after the cryogen was exhausted in September 2010, only two filters with central wavelengths of 3.4 and 4.6 $\mu$m, named W1 and W2 respectively, were used.
The 5$\sigma$ detection limits are $W1=15.8$ and $W2=14.4$ on a single-exposure image with a typical exposure time of $\sim$9 seconds.
If interested in events with variability time scales of months to years and careless about any intra-day variability, one can obtain time-resolved coadded (TRC) images by stacking the single-exposure images taken during each WISE visit \citep{Meisner2018}.
The detection limits on the TRC images are $\sim$1.3 magnitudes fainter than on the single-exposure images.
Compared with Spitzer, WISE is inferior in spatial resolution (6.1$\arcsec$ and 6.6$\arcsec$ in the W1 and W2 bands) and depth, but superior in sample size and monitoring duration.
Thus WISE data is suitable for the study of rare and bright SNe, such as SLSNe.

In this work, we studied the MIR emission of a sample of nearby SLSNe occurred in recent years using WISE data.
Throughout the paper, we adopted cosmological parameters of $H_0=69.6$ km s$^{-1}$ Mpc$^{-1}$, $\Omega_m=0.286$ and $\Omega_\Gamma=0.714$ from \cite{Planck2016}.


\section{Sample Selection and MIR photometries}
\label{sec:data_reduction}

\subsection{Sample Selection}
\label{sec:sample}

We selected a sample of nearby SLSNe from Open Supernova Catalog \citep[OSC,][]{Guillochon2017} and the Weizmann Interactive Supernova Data Repository \citep[WISeREP,][]{Yaron2012} with four selection criteria as follows.
Firstly, the SN has a peak absolute magnitude $<-21$ \citep{Gal-Yam2012}.
Secondly, the SLSN exploded between 2015 and 2018, so the NEOWISE-R observation spans from the explosion to years after the optical peak.
Thirdly, the SLSN has a redshift $<$0.12, and hence the 3$\sigma$ detection limits on WISE TRC images (W1$\sim$17.6 mag and W2$\sim$16.2 mag) correspond to monochromatic luminosities ($\lambda L_\lambda$) at 3.4 and 4.6 $\mu$m of $\lesssim10^{8.5}$ $L_\odot$.
Finally, the SLSN was spectroscopically classified with certainty.
Note that we not only required a distinction between SLSN-I and SLSN-II, but also required that it was not considered to be a tidal disruption event (TDE), and hence SN 2016ezh \citep[$=$PS16dtm,][]{Blanchard2017} was excluded.

The final sample includes 11 SLSNe.
We collected the basic information of the SLSNe, including the spectral type, the time of optical peak ($t_{\rm peak}$), the redshift, and the position, from the literature.
We did not find the $t_{\rm peak}$ for LSQ15abl and SN 2017err from the literature, thus we estimated the $t_{\rm peak}$ for LSQ15abl using archival data and obtained the $t_{\rm peak}$ for SN 2017err via private communication (see details in Appendix A).
The information is listed in Tab.~\ref{tab1}.
The sample includes 8 SLSN-I and 3 SLSN-II.
Note that SN2017ens, which we have classified as SLSN-I, is special in that its spectrum was type I around the peak and became type IIn in the late time \citep{Chen2018}.
Throughout this paper, we calculated the phases using the time of optical peak as the zero point, because we lack information on the time of explosion for most SLSNe in the sample.

\begin{table*}
\footnotesize
\caption{Information of the SLSNe sample.}
\begin{threeparttable}
\begin{tabular}{ccccccc}
\hline
Name       & Type    & optical peak & z     & RA & DEC & Ref\\
\hline
PS15br     & SLSN-II  & 2015 Mar 8 (57089.3)  & 0.101  & 11:25:19.22 & $+$08:14:18.9 & 1 \\
SN 2015bn  & SLSN-I   & 2015 Mar 21 (57102.0) & 0.1136 & 11:33:41.55 & $+$00:43:33.4 & 2 \\
LSQ15abl   & SLSN-II  & 2015 Apr 24 (57136)   & 0.087  & 09:40:29.50 & $-$04:11:32.3 & 3 \\
SN 2016eay & SLSN-I   & 2016 Jun 2 (57541.4)  & 0.1013 & 12:02:51.71 & $+$44:15:27.4 & 4,5 \\
SN 2017ens & SLSN-I   & 2017 Jun 20 (57924.0) & 0.1086 & 12:04:09.37 & $-$01:55:52.2 & 6 \\
SN 2017egm & SLSN-I   & 2017 Jun 21 (57925.8) & 0.0307 & 10:19:05.62 & $+$46:27:14.1 & 7 \\
SN 2017err & SLSN-IIn & 2017 Jul 1 (57935)    & 0.107  & 11:11:25.10 & $+$00:06:58.5 & 8 \\
SN 2017gci & SLSN-I   & 2017 Aug 25 (57990.3) & 0.0873 & 06:46:45.02 & $-$27:14:55.8 & 9 \\
SN 2018bgv & SLSN-I   & 2018 May 15 (58253.2) & 0.0795 & 11:02:30.29 & $+$55:35:55.8 & 10 \\
SN 2018bsz & SLSN-I   & 2018 May 29 (58267.5) & 0.0267 & 16:09:39.11 & $-$32:03:45.6 & 11 \\
SN 2018hti & SLSN-I   & 2018 Dec 12 (58464.6) & 0.0612 & 03:40:53.75 & $+$11:46:37.3 & 12 \\
\hline
\end{tabular}
\begin{tablenotes}
   \item The references are: (1) \cite{Inserra2018}; (2) \cite{Nicholl2016}; (3) \cite{Prajs2015}; (4) \cite{Yan2017}; (5) \cite{Nicholl2017}; (6) \cite{Chen2018}; (7) \cite{Bose2018}; (6) \cite{Chen2017}; (9) \cite{Fiore2021}; (10) \cite{Lunnan2020}; (11) \cite{Anderson2018}; (12) \cite{Lin2020}.
\end{tablenotes}
\end{threeparttable}
\label{tab1}
\end{table*}

\subsection{MIR Photometries}
\label{sec:mirphoto}

We made MIR photometries of the SLSNe using the WISE TRC images produced by \cite{Meisner2018}.
There are 16 or 17 available images of each SLSN position in each of the W1 and W2 bands: the first two or three were taken between 2010 and 2011, and the remaining 14 were taken between 2014 and 2020.
For each SLSN, we created a reference image in each band by combining the pre-explosion images using SWarp \citep{Bertin2010}.
Here the pre-explosion images refer to images taken at least 60 days before the optical peak, and there are 4--12 pre-explosion images for each SLSN.
For all the post-explosion images, we performed image subtraction with ISIS \citep[version 2.2,][]{Alard1998,Alard2000}.
We display the pre-explosion, post-explosion, and difference images in Appendix B.

We performed PSF-fitting photometry to the SLSNe on the difference images using a method described in \cite{Lang2016}.
When fitting with PSF, the positions of the targets were fixed on the SLSN positions from optical surveys.
For most of the SLSNe, the host galaxy is weak and we measured the error through standard procedures.
The host galaxies of LSQ15abl, SN 2017egm, and SN 2018bsz are bright and introduce an extra error.
The extra error can lead to false detections on reference-subtracted pre-explosion images.
Therefore we estimated the extra error using the root square mean of the fluxes of the false detections and added it to the total error.
We list all the photometries in Appendix B.
We set a 3$\sigma$ threshold for a detection and calculated a 3$\sigma$ upper limit when there is no detection.

SN 2018bsz, an SLSNe in the sample, was also observed by Spitzer in I1 and I2 bands.
We found that the fluxes of SN 2018bsz measured by WISE and Spitzer are in agreement, as can be seen in detail in Appendix C.
This supports the reliability of studying SLSNe using WISE data.

\section{Data Analysis and Results}
\label{sec:analysis_result}

\subsection{MIR Light Curves}
\label{sec:mirlc}

MIR emission was detected from 10 out of 11 SLSNe, and the only exception is SN 2018bgv.
Fig. 1 shows the light curves in W1 and W2 bands (expressed as $\lambda L_\lambda$) of the 10 SLSNe.
SN 2017egm, SN 2018bsz, and SN 2018hti, the three closest SLSNe ($z<0.07$) in the sample, show a similar pattern in the MIR LCs.
Their MIR emission peaks between 0d and $+$100d with a monochromic luminosity of $\sim10^{8}\ L_\odot$,
then decreases continuously, and finally fades between $+$400d and $+$500d.
Limited by large measurement errors, we cannot clearly tell the shape of the LCs of SN 2015bn, LSQ15abl, SN 2016eay, and SN 2017gci.
Nonetheless, an assumption that they also have the same LC shape as the three closer SLSNe does not contradict the observations.
Unlike them, PS15br's MIR emission rebrightens at $+$404d after an initial decline between $+$78d and $+$253d.
For SN 2017ens, although an initial decline was not recorded likely due to incomplete sampling, a late-time rebrightening can be clearly seen between $+$169d and $+$497d.
SN 2017err's W1-band fluxes on $+289$d and $+477$d show 2.3$\sigma$ and 2.1$\sigma$ increases compared with the flux on $+147$d.
Combining the two, there is a 99.8\% probability that a rebrightening is also seen in SN 2017err.
The MIR rebrightening in the three SLSNe lasts at least one to two years.
During the rebrightening, the MIR emission reach a peak monochromic luminosity of $\sim10^{8.5}\ L_\odot$ in both the two bands.

\begin{figure}
 \centering{
  \includegraphics[scale=0.68]{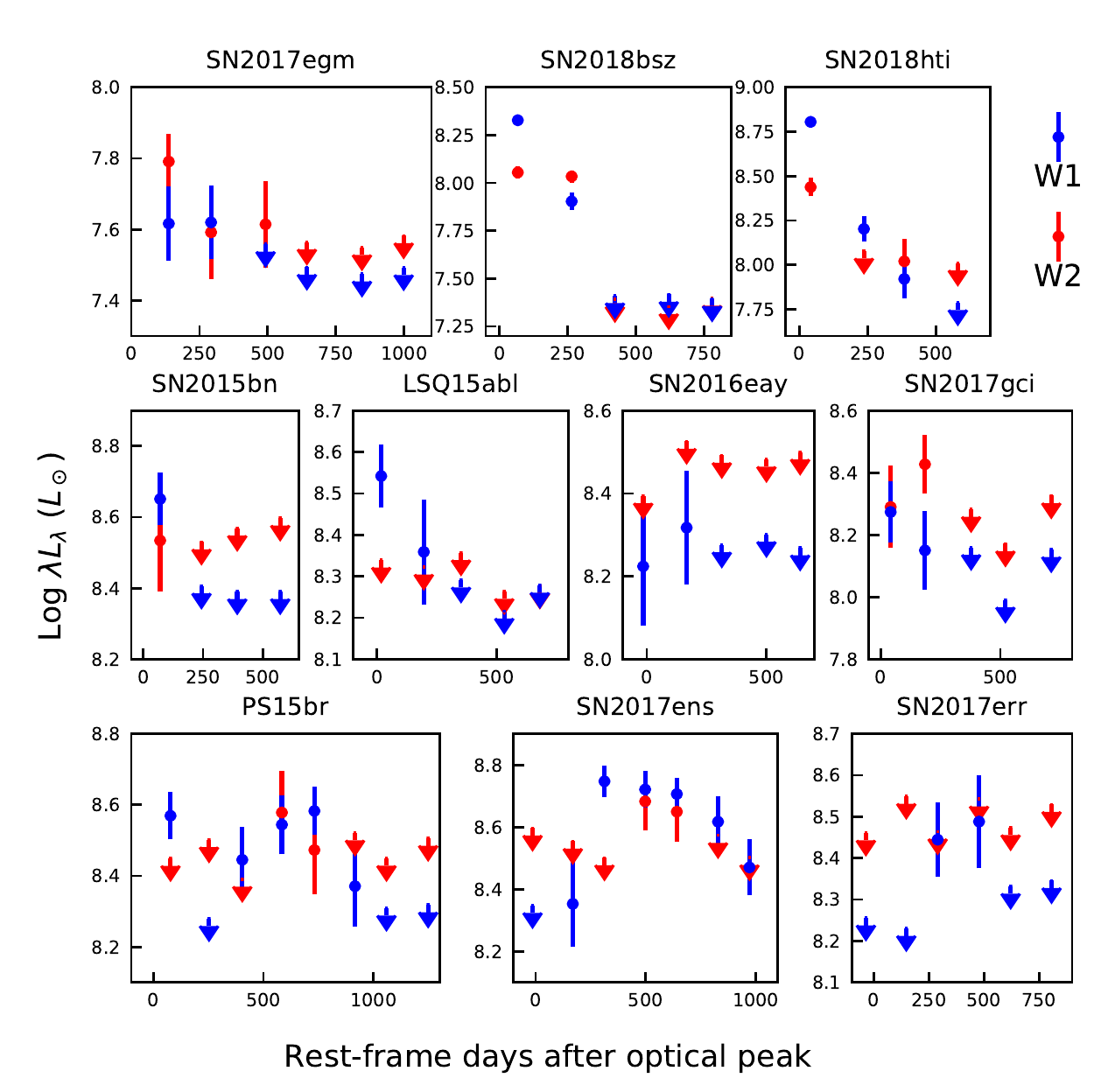}
  \caption{The MIR light curves of the 10 SLSNe with WISE detections.
  The luminosities at 3.4 and 4.6 $\mu$m are shown in blue and red, respectively.}}
 \label{fig:mirlc}
\end{figure}

\subsection{MIR Excess and Dust Properties}
\label{sec:mirexcess}

The early-time MIR emission in normal SNe can be the counterpart of the optical blackbody.
While the late-time MIR emission, if present, usually exceeds the expected flux level inferred from optical data, and is likely the thermal emission of dust \citep[e.g.,][]{Tinyanont2016,Szalai2019}.
The situation is likely similar in SLSNe.
In this section, we are concerned with the MIR excess, i.e. the dust emission.

For SN 2015bn, LSQ15abl, and SN 2016eay, the MIR detections are at low levels possibly due to their greater distance ($z\sim0.1$), and we did not analyse them in detail.
We divided the remaining 7 SLSNe into two groups according to the presence of the rebrightening feature.
We refer to the two groups as ``declining group'' and ``rebrightening group'' hereafter.
The declining group includes SN 2017egm, SN 2017gci, SN 2018bsz, and SN 2018hti, and the rebrightening group includes PS15br, SN 2017ens, and SN 2017err.

\subsubsection{The Declining Group}
\label{sec:mirexcess_decline}

A red MIR color is a probe of dust emission among SNe.
We show the variation of the WISE color $W1-W2$ for the declining group in Fig. 2. 
The MIR colors of SN 2018bsz and SN 2018hti clearly turn red over time, and in the late time (after $+$200d or $+$300d) their colors ($W1-W2\sim1.2$) correspond to blackbody temperatures ($T_{\rm BB}$) of about 1000 K.
The variation suggests that although their early-time MIR emission can be dominated by the long-wavelength tail of an optical/NIR blackbody, there must be significant dust emission in the late time.
For example, assuming the spectrum is the sum of two blackbodies with $T_{\rm BB}$ of 6000 and 600 K, a color of $W1-W2\sim1$ indicates that 70\% of the W2-band emission is from dust\footnote{$z=0.05$ is assumed.}.
For typical dust spectra, we calculated a conversion between $W1-W2$ color and Spitzer/IRAC's $I1-I2$ color (see Appendix D for details) as:
\begin{equation}
I1-I2 \approx 0.77 (W1-W2) - 0.02
\end{equation}
With this conversion, $W1-W2\sim1.2$ corresponds to $I1-I2\sim0.9$, consistent with those of normal SNe with dust emission detected \citep{Szalai2021}.
We detected no significant changes in the MIR colors of SN 2017egm and SN 2017gci, and the red colors ($W1-W2\sim1$) suggest that dust emission is also notable in them.
Using data points with red colors ($W1-W2>0.8$) and assuming that $50\%\sim100\%$ of the W2-band emission comes from dust, we estimated that the W2-band monochromatic luminosity of the dust emission is $10^7\sim10^8$ $L_\odot$.

\begin{figure}
\centering{
  \includegraphics[scale=0.8]{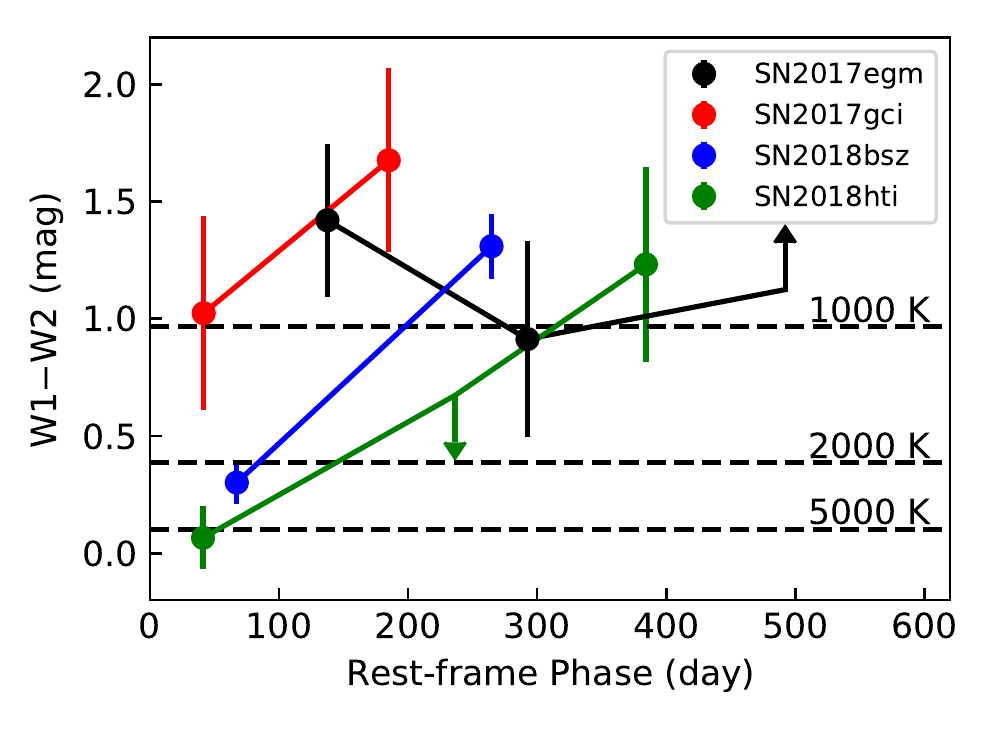}
  \caption{The variation of the MIR color W1$-$W2 for the 4 SLSNe in the dropping group.
  We show the MIR colors of blackbodies with $T_{\rm BB}$ of 1000, 2000, and 5000 K at $z=0.05$ for a comparison.}}
  \label{fig:mirexcess1}
\end{figure}

The dust emission in SN 2018bsz was well studied by \cite{Chen2022} combining with ground-based optical/NIR data as well as Spitzer MIR data.
They concluded that the dust mass was between $10^{-4}$ and $10^{-2}$ $M_\odot$.
For the other three, although we cannot make a similar analysis due to a lack of data, we speculated that the dust mass is also between $10^{-4}$ and $10^{-2}$ $M_\odot$ given that the three are similar to 2018bsz in MIR luminosity and color.

\subsubsection{The Rebrightening Group}
\label{sec:mirexcess_rebright}

\begin{figure}
\centering{
  \includegraphics[scale=0.75]{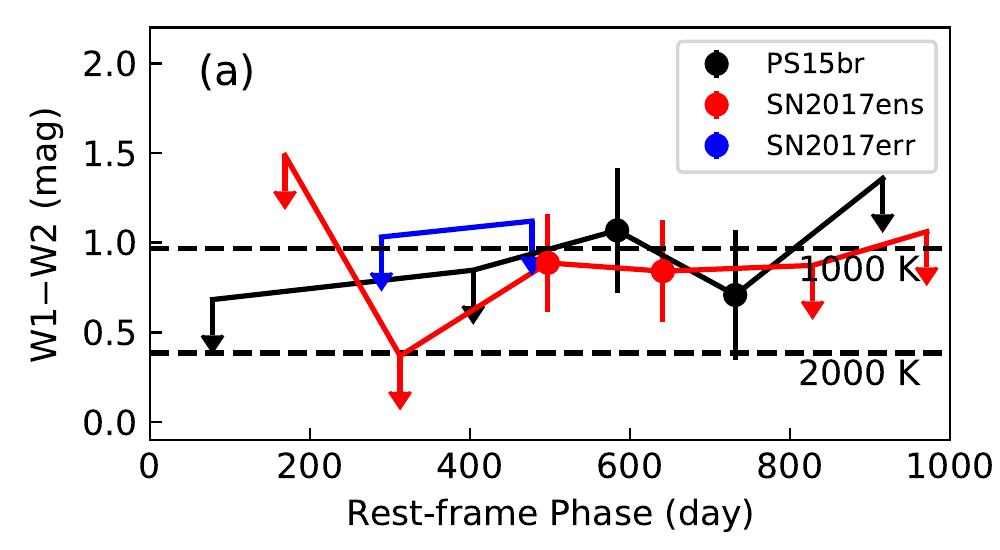}
  \includegraphics[scale=0.75]{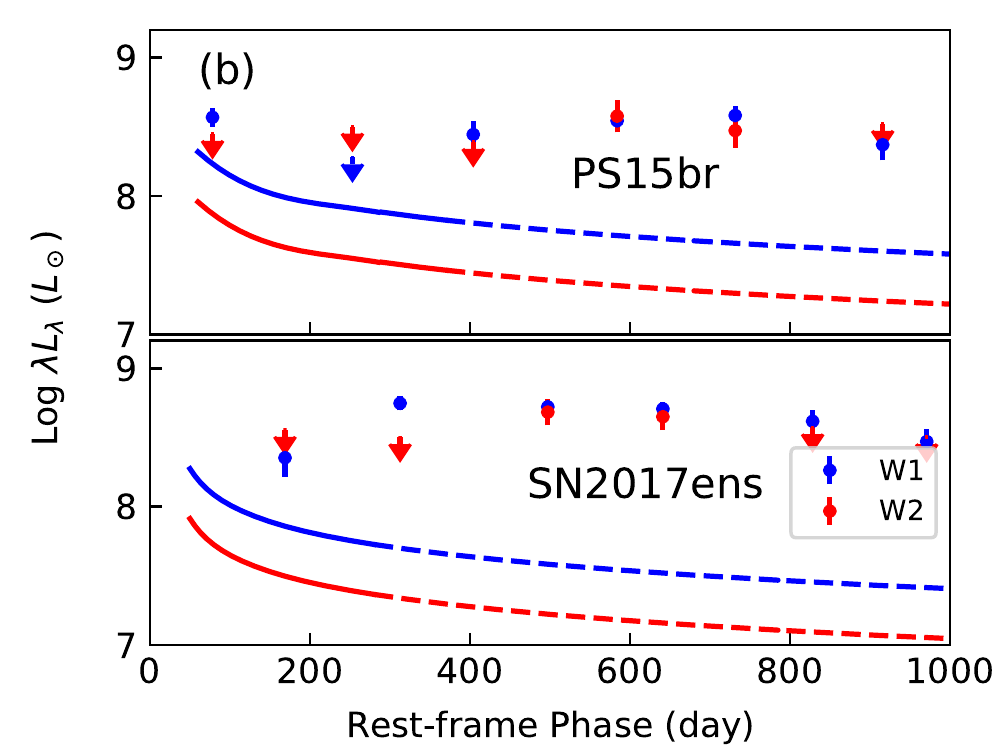}
  \caption{
    {\bf (a)}: The variation of the MIR color W1$-$W2 for the 3 SLSNe in the rebrightening group.
    We show the MIR colors of blackbodies with $T_{\rm BB}$ of 1000 and 2000 K at $z=0.1$ for a comparison.
    {\bf (b)}: The model LCs of the photosphere in W1 and W2 bands for PS15br and SN 2017ens, compared with the observed MIR data.
    We show the predictions by optical/NIR data with solid lines, and further extrapolations with dashed lines.}}
  \label{fig:mirexcess2}
\end{figure}

PS15br, SN 2017ens, and SN 2017err show rebrightenings with luminosities of $>10^8$ $L_\odot$ $1\sim2$ years after the optical peak.
A late-time MIR rebrightening among SNe is generally caused by delayed dust emission, and the same likely goes for the three SLSNe.
As can be seen in Fig. 3(a), in the late time the MIR colors of the three SLSNe are $W1-W2\sim1$, consistent with dust emission.
We further estimated the contribution to the MIR emission by the SLSN photosphere in PS15br and SN 2017ens, which have abundant monitoring data in optical and NIR bands.
We constructed the bolometric LCs for PS15br and SN 2017ens with methods described in detail in Appendix E.
For both the two SLSNe, the temperature of the optical blackbody becomes stable to around 6000 K after $+$60d \citep{Inserra2018,Chen2018}.
Thus we predicted the MIR flux of the SLSN photosphere by assuming a blackbody spectrum with a temperature of 6000 K.
The results, displayed in Fig. 3(b), show that the SLSN photosphere contributes less than 15\% to the W1-band flux, and less than 10\% to the W2-band flux.
Therefore the majority of MIR emission comes from dust during the rebrightening.

\begin{table*}
\footnotesize
\caption{Dust parameters for PS15br and SN 2017ens}
\begin{tabular}{cccccccc}
\hline
SLSN & Phase & \multicolumn{2}{c}{Graphite} & \multicolumn{2}{c}{Amorphous Carbon} & \multicolumn{2}{c}{Silicate} \\
     &       & $t_d$ & $M_d$            & $t_d$ & $M_d$            & $t_d$ & $M_d$\\
     & day   & K     & $10^{-2}M_\odot$ & K     & $10^{-2}M_\odot$ & K     & $10^{-2}M_\odot$\\
\hline
PS15br    &584 & $670\pm120$ & $1.8^{+3.8}_{-1.2}$ & $780\pm160$ & $1.6^{+3.5}_{-1.1}$ & $840\pm190$ & $3.6^{+7.9}_{-3.0}$\\
          &732 & $810\pm170$ & $0.6^{+1.3}_{-0.4}$ & $990\pm260$ & $0.5^{+1.1}_{-0.3}$ & $1080\pm320$& $1.1^{+2.6}_{-0.9}$\\
\hline
2017ens   &497 & $740\pm110$ & $1.4^{+2.0}_{-0.8}$ & $880\pm160$ & $1.2^{+1.9}_{-0.7}$ & $950\pm190$ & $2.8^{+4.2}_{-2.2}$\\
          &641 & $760\pm110$ & $1.2^{+1.7}_{-0.7}$ & $910\pm170$ & $1.0^{+1.5}_{-0.6}$ & $980\pm200$ & $2.3^{+3.5}_{-1.8}$\\
\hline
\end{tabular}
\label{tab2}
\end{table*}

\begin{figure}
\centering{
  \includegraphics[scale=0.8]{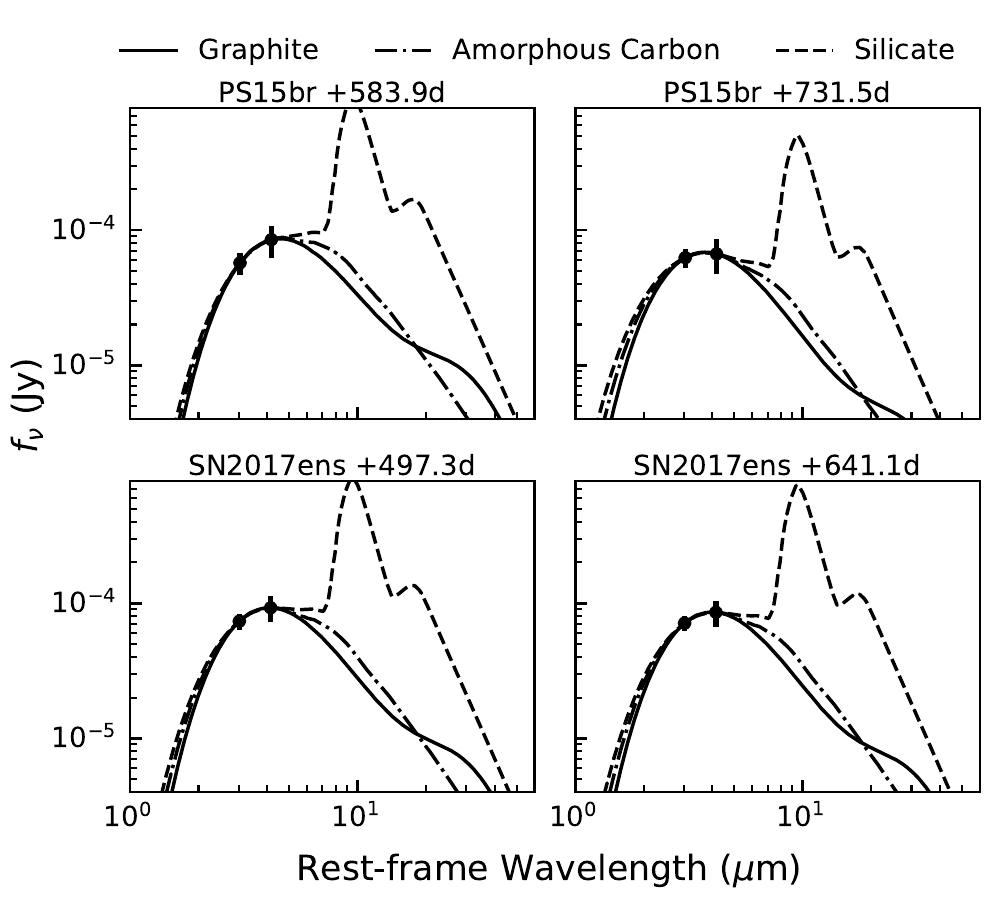}
  \caption{The SED fitting results for PS15br and SN 2017ens.}}
  \label{fig:dust_sed}
\end{figure}

We measured the properties of the dust that causes the MIR rebrightening in PS15br and SN 2017ens using WISE observations showing detections in both the two bands.
These include $+584$d and $+732$d observations for PS15br, and $+497$d and $+641$d observations for SN 2017ens.
We fit the observed SED using models of dust emission.
We assumed optically thin dust whose spectrum is expressed as \citep[e.g.,][]{Hildebrand1983,Fox2010}:
\begin{equation}
L_\nu = 4 \pi M_d B_\nu(T_d) \kappa_\nu(a),
\end{equation}
where $M_d$ is the dust mass, $T_d$ is the dust temperature, $B_\nu$ is the Planck function, $a$ is the grain radius, and $\kappa_\nu$ is the dust mass absorption coefficient.
We tried dust compositions of graphite, amorphous carbon, and silicate\footnote{For graphite and silicate, we use the absorption coefficients from \cite{Fox2010}. And for amorphous carbon, we use those from \cite{Rouleau1991}.}, and adopted a grain radius of 0.1 $\mu$m following previous works \citep[e.g.,][]{Tinyanont2016,Szalai2019}.
We show the models that match the observations in Fig. 4, and list the dust parameters in Tab.~\ref{tab2}.
We found that $T_d$ are between 600 and 1100 K, and $M_d$ are $(0.5\sim1.8)\times10^{-2}$ $M_\odot$ for graphite or amorphous carbon dust, and are $(1\sim4)\times10^{-2}$ $M_\odot$ for silicate dust.
By integrating the best-fitting spectral model, we measured dust luminosities $L_{\rm dust}$ of $10^{8.6}\sim10^{9.1}$ $L_\odot$.

\begin{figure}
\centering{
  \includegraphics[scale=0.8]{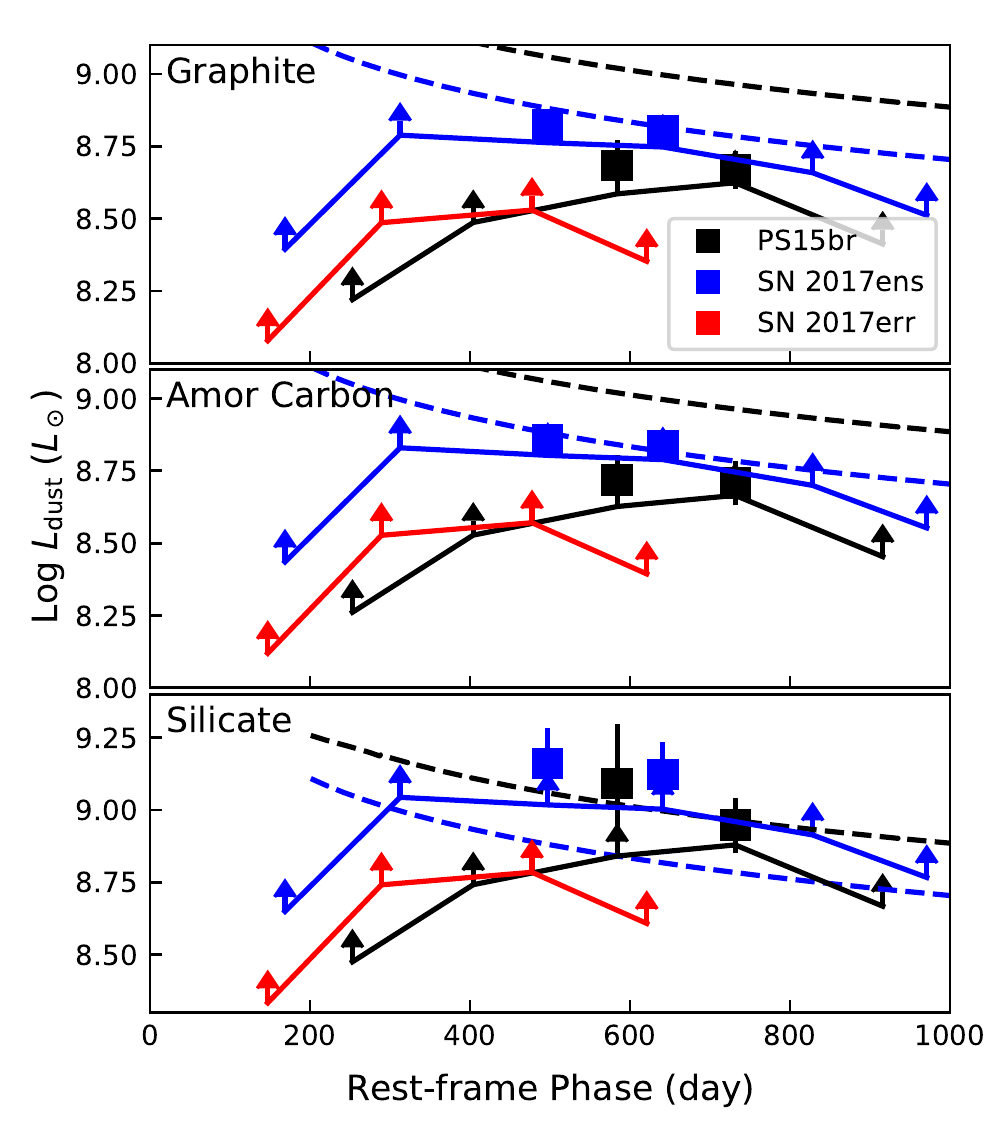}
  \caption{The $L_{\rm dust}$ curves for PS15br, SN 2017ens, and SN 2017err assuming different dust compositions.
  The squares are from spectral fitting, and the upper limits are converted from $\lambda L_\lambda$ in the W1 band.
  We also show the inferred bolometric LCs for PS15br and SN 2017ens for a comparison.}}
  \label{fig:dust_lum}
\end{figure}

However, for most epochs of WISE observations, there is no significant detection in the W2 band, preventing us from calculating $L_{\rm dust}$ via a similar fitting.
Nonetheless, there is a minimum integrated luminosity for SED models that are consistent with the only data point in the W1 band.
This minimum value sets the lower limit of the $L_{\rm dust}$.
Considering bolometric correction $C_B$ in the W1 band expressed as:
\begin{equation}
C_B = \frac{L_{\rm dust}} {\lambda L_\lambda(W1)}.
\end{equation}
For given dust properties and a given redshift, $C_B$ is a function of the dust temperature $T_d$, as can be seen in detail in Appendix F.
The function has a minimum value ${\rm min}(C_B)$ of 1.10, 1.21, and 1.98 for graphite, amorphous carbon, and silicate dust, respectively, at $z=0.105$.
Thus we obtain:
\begin{equation}
L_{\rm dust}>{\rm min}(C_B) \times \lambda L_\lambda(W1).
\end{equation}

In Fig. 5, we show the $L_{\rm dust}$ curves assuming different dust compositions.
The dust luminosities are in the range of $10^8\sim10^9$ $L_\odot$.

\begin{table}
\footnotesize
\caption{Total Energy of Dust Emission}
\begin{tabular}{cccc}
\hline
SLSN & \multicolumn{3}{c}{$E_{\rm dust}$ ($10^{50}$ erg)} \\
     & Graphite & Amor Carbon & Silicate \\
\hline
PS15br       &$>0.84$ & $>0.91$ & $>1.73$ \\
SN 2017ens   &$>1.43$ & $>1.56$ & $>2.79$ \\
SN 2017err   &$>0.48$ & $>0.53$ & $>0.87$ \\
\hline
\end{tabular}
\label{tab3}
\end{table}

We estimated the total energy of the dust emission as:
\begin{equation}
E_{\rm dust} = \sum L_{\rm dust} \times \Delta t,
\end{equation}
where $\Delta t$ is the time interval of WISE observations in the rest frame.
The above summation includes all late-time WISE observations as long as there are $>2\sigma$ detections in the W1 band.
For most epochs, there are only lower limits of $L_{\rm dust}$, so we calculated the lower limit of $E_{\rm dust}$.
The results are listed in Tab.~\ref{tab3}.
For all the three SLSNe, the total energies are on the order of $10^{50}$ erg.

\section{Dust Origin and Heating Mechanism}
\label{sec:dust_origin}

\subsection{Time-delay Analysis}
\label{sec:dust_timedelay}

The MIR emitting dust around SNe can be pre-existing or newly formed.
Pre-existing dust is sufficient to produce significant emissions only outside the dust-free cavity formed by the SNe.
The radius of the dust-free cavity, known as evaporation radius $r_{\rm evap}$, is determined by the SN peak luminosity $L_{\rm peak}$ \citep[e.g.,][]{Dwek1985,Fox2011} as:
\begin{equation}
r_{\rm evap} = \left( \frac{ L_{\rm peak} }{ 16 \pi \sigma T_{\rm evap}^4 \langle Q \rangle(T_{\rm evap}) } \right)^{1/2},
\end{equation}
where $\sigma$ is the Stefan-Boltzmann constant, $T_{\rm evap}$ is the vaporization temperature of the dust and $\langle Q \rangle(T_{\rm evap})$ is the Planck-averaged value of the dust emissivity at the vaporization temperature.
The emission from pre-existing dust is therefore delayed relative to the optical emission, and the time delay $\tau_{\rm IR}$, which is defined as the time relative to the optical peak, depends on the heating mechanism.
If the dust is heated by the SN peak luminosity (IR echo), then $\tau_{\rm IR}\sim r/c$, where $r$ is the distance of the emitting dust, and $c$ is the speed of light.
Simulations \citep[e.g.,][]{Dwek1983,Dwek1985} show that under typical conditions, the delay for 3--5 $\mu$m emission is about $1\sim10$ times $r_{\rm evap}/c$.
If the dust is collisionally heated by shock, $\tau_{\rm IR}\sim r/v_{\rm shock}-t_{\rm rise}$, where $v_{\rm shock}$ is the shock velocity, and $t_{\rm rise}$ is the time from the explosion to the optical peak.
The shock velocity can be approximated by the photospheric velocity near the peak, which is between 6000 and 20,000 km s$^{-1}$ for most SLSNe with a median value of 12,000 km s$^{-1}$, while the rise time of SLSNe is between 20 and 100 days with a median value of $\sim$40 days \citep[e.g.,][]{Quimby2018,Inserra2018,Konyves-Toth2021}.
There is also a time delay in the emission from newly formed dust because it takes time for the dust to form.
For common core-collapse SNe (CCSNe), the MIR emission from newly formed dust becomes significant after several hundred days \citep[e.g.,][]{Sugerman2006,Szalai2021}.

In Fig. 6, we display the $\tau_{\rm IR}$ and $L_{\rm peak}$\footnote{The $L_{\rm peak}$ value of SN 2017err was calculated by us ($1.1\times10^{44}$ erg s$^{-1}$, see Appendix A), and those of the other SLSNe were collected from the literature.} for the seven SLSNe with dust emission detected.
We also show the constraint functions for different dust models for comparison.
When calculating $r_{\rm evap}$, one should note that the pre-existing dust may undergo twice radiative heating, one by the shock breakout and the other by the optical peak.
Since the luminosity and duration of shock breakouts in SLSNe are loosely constrained, we estimated the lower limit of the $r_{\rm evap}$ by considering only the optical peak.
We assumed graphite dust, and set grain size $a=0.1$ $\mu$m and $T_{\rm evap}=1900$ K following \cite{Dwek1985}.
Under these assumptions, $r_{\rm evap}=0.11 \sqrt{L_{44}}$ light-year, where $L_{44}$ is the $L_{\rm peak}$ in unit of $10^{44}$ erg s$^{-1}$.
For the IR echo case, we show the region with $r_{\rm evap}/c < \tau_{\rm IR} < 10 r_{\rm evep}/c$.
For the shock-heating case, we show the region with $\tau_{\rm IR}>r_{\rm evap}/v_{\rm shock}-t_{\rm rise}$ by assuming a $t_{\rm rise}$ of 40 days and a $v_{\rm shock}$ of 12000 km s$^{-1}$.
For the case of newly formed dust, we simply show a region between 100 and 1000 days.
The $\tau_{\rm IR}$ given by the shock-heating model is above 1000 days, and may be longer if considering the destruction of the pre-existing dust by the shock breakout.
This is inconsistent with the observation, and therefore it is unlikely that the MIR emission of SLSNe comes from pre-existing dust heated by shock.
It is difficult to draw conclusions on the IR echo and newly formed dust models simply based on time delay.
We will discuss them in detail next.

\begin{figure}
\centering{
  \includegraphics[scale=0.8]{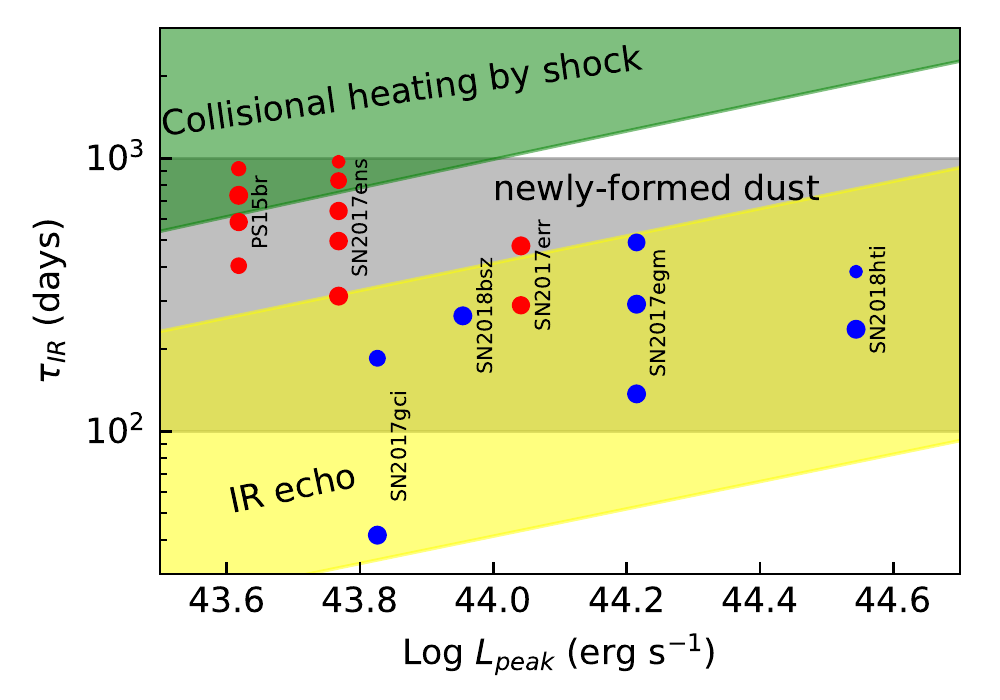}
  \caption{The $\tau_{\rm IR}$ and $L_{\rm peak}$ for the seven SLSNe analysed in detail, and the constraint functions for dust models.
  The data for SLSNe in the dropping group and the rebrightening group are labeled using blue and red circles, respectively.
  The size of the circles represents the W1-band magnitude (larger for brighter), which is normalized using the peak value.}}
  \label{fig:irdelay}
\end{figure}

\subsection{IR Echo?}
\label{sec:dust_irecho}

We simulated the IR echo of an SLSN and see what its MIR LC should look like.
We ran the simulation following \cite{Dwek1983,Dwek1985} with the assumptions listed as followings.
The SLSN has a peak luminosity $L_{\rm peak}=10^{44}$ erg s$^{-1}$ and drops exponentially after the peak with an e-folding time of $t_{\rm SN}=50$ day.
The dust is distributed spherically symmetrically around the SLSN with an outer radius $R_2=3$ light-year.
The dust is composed of graphite grains with a uniform radius of $a=0.1$ $\mu$m.
The peak SN luminosity causes a new spherical inner surface of dust with a radius $r_{\rm in}=r_{\rm evap}=0.11$ light-year.
The grain number density is inversely proportional to the square of the distance ($n_d\propto r^{-2}$), and the normalization is set so that the V-band attenuation $A_V$ is 0.05.
We display the W1-band LC from the simulation in Fig. 7.
We found that the IR echo peaks with a time delay, which can be roughly expressed as $\tau_{\rm IR}\sim3 r_{\rm evap}/c$ when the dust is optically thin in the UV/optical bands, and is about 100 days for typical parameters.
The peak W1-band luminosity is roughly proportional to $n_d$ for a given grain radius, and is about $10^8$ $L_\odot$ under our assumptions.
In addition, the model predicts that the MIR LC keeps decreasing after taking into account the photospheric emission.

\begin{figure}
\centering{
  \includegraphics[scale=0.8]{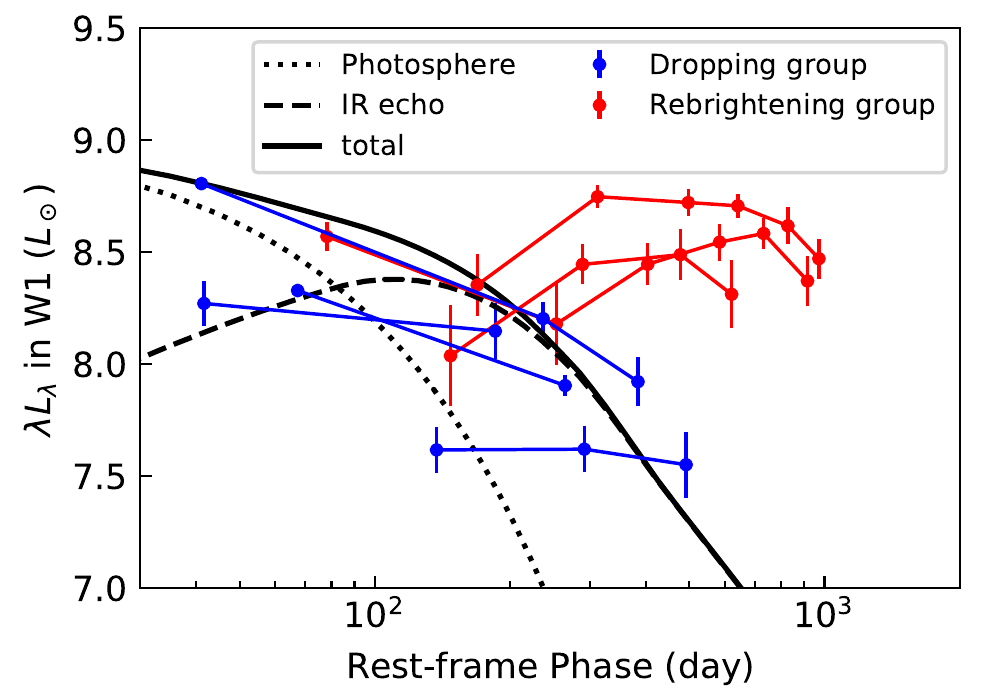}
  \caption{The data and simulated LC in the W1-band for the seven SLSNe analysed in detail.}}
  \label{fig:echo_model1}
\end{figure}

For the four SLSNe in the dropping group, the observed MIR LC (blue points in Fig. 7) roughly agrees with the IR echo model in terms of time delay and luminosity.
The inconsistency with the curve shown in Fig. 7 can be reconciled by adjusting the grain radius $a$ and the number density $n_d$.
The model predicts that the IR echo dominates the MIR emission after $+100$d, causing the MIR color to turn red.
This is also consistent with the observations (see Fig. 2).
Assuming that the echoing dust originates from the mass loss of the progenitor star, the $n_d$ given by the model corresponds to a mass loss rate $\dot{M}$ of $10^{-5}\sim10^{-4} (f_d/0.01) (v_{\rm loss}/{\rm 100\ km\ s^{-1}})$ $M_\odot$ yr$^{-1}$, where $f_d$ is the mass fraction of the dust, and $v_{\rm loss}$ is the velocity of the mass loss.
This rate is reasonable for high mass stars, which are possible SLSN progenitors.
Thus the presence of echoing dust is natural.

However, for the three SLSNe in the rebrightening group, the above IR echo model can not reproduce the late time emission for two reasons.
One reason is that the model predicts a peak before $+$200d while the observed LC peaks at 300--600 days.
The second reason is that the predicted dust luminosity is not high enough when the dust is optically thin to the SLSN UV/optical emission ($A_V<0.1$).
Fitting the model to observations requires larger $\dot{M}$ and larger $r_{\rm in}$.
Potential SN progenitors that have mass loss rate $\dot{M}>10^{-4}\ M_\odot\ {\rm yr}^{-1}$ include luminous blue variable (LBV) stars \citep[e.g. $\eta$ Car,][]{Smith2003}, extreme red supergiant stars \citep[e.g. VY CMa,][]{Decin2006}, and dusty Wolf-Rayet star \citep[e.g. WR 48a,][]{Zhekov2014}.
In addition, the dust around the LBV stars can be rich in silicate \citep[e.g.,][]{Morris2017}, which has a larger $r_{\rm evap}$ than graphite and may explain the large $r_{\rm in}$ required by the observations.
Thus, using the IR echo model to explain the late-time MIR emission of the three SLSNe requires that the progenitor stars have extreme mass losses, and the dust is rich in silicates.

We tested the possibility by examining the energy balance and the required echoing dust.
In section 3.2.2, we estimated the lower limit of the total energy of dust emission $E_{\rm dust}$ for the three SLSNe assuming silicate dust.
The energies are at least $1.7\times$, $2.8\times$ and $0.9\times10^{50}$ erg for PS15br, SN 2017ens and SN 2017err, respectively.
However, the total energies released by PS15br and SN 2017ens near their peaks ($<$200 days after peak) are only about $3\times10^{50}$ erg.
Thus the IR echo model requires that almost all of the UV/optical emission near the peak is converted to IR emission by warm dust.
In addition, the observed IR time delay requires that the warm dust is on the light-year scale.
We show that these requirements can hardly be satisfied for the following three reasons.
Firstly, the dust does not cover all directions because little dust extinction is observed in the light of sight, and hence some UV/optical emissions can escape.
Secondly, part of scattered UV/optical emission and short-wavelength dust emission is absorbed with a large time delay, and the dust heated by them is likely too cold to generate significant 3 $\mu$m emission.
Thirdly, assuming a typical grain radius distribution, for dust on a light-year scale to obscure optical emission, the required dust mass is at least $\sim1$ $M_\odot$, which is too large even for $\eta$ Car analogues.
Therefore, IR echo cannot account for late-time MIR emission from PS15br and SN 2017ens.
The peak luminosity of SN 2017err ($1.1\times10^{44}$ erg s$^{-1}$) is about twice those of PS15br and SN 2017ens (4.15 and 5.86 $\times10^{43}$ erg s$^{-1}$, respectively).
Assuming the bolometric LC of SN 2017err has the same shape as that of SN 2017ens, the near-peak UV/optical emission of SN 2017err is about $6\times10^{50}$ erg, enough to power the observed dust emission with the energy of $1.2\times10^{50}$ erg.
Thus an IR echo model may apply to SN 2017err, although it is difficult to be fully tested due to insufficient data.

\subsection{Echo of Shock Emission from CSM Interaction?}
\label{sec:dust_shockecho}

We found that the three SLSNe in the rebrightening group, including PS15br, SN 2017ens, and SN 2017err, all exhibit features of strong CSM interaction.
PS15br shows a strong, multi-component H$\alpha$ emission after $+$200d, which was interpreted as an interaction of the ejecta with an asymmetric CSM \citep{Inserra2018}.
In the late-time spectrum of SN 2017ens, Balmer emission lines consist of components with widths of $\sim$2000 and several tens km s$^{-1}$, possibly representing postshock and undisturbed CSM, respectively \citep{Chen2018}.
Also, strong coronal lines are present, possibly powered by the interaction.
For SN 2017err, interaction is also suggested by the narrow emission lines in the near-peak spectrum, which leads to an SLSN-IIn classification by Chen et al. (2017).
These features of strong CSM interactions lead us to a conjecture: does the late-time MIR emission come from the pre-existing dust which is radiatively heated by shock emission, as supposed by \cite{Fox2010,Fox2011,Fox2013} for SN 2005ip and other SNe IIn?

We examined whether this model applies to PS15br and SN 2017ens.
As can be seen from Fig. 5, the dust luminosity in the late time is close to the predicted power of shock emission.
This requires thick asymmetric dust at a scale of several times $r_{\rm evap}$ (0.1--0.3 light-year), which absorbs most of the shock emission except for that escaping from the line of sight and converts it to IR emission.
If such dust exists, it must also produce an IR echo in response to the peak SN luminosity.
Based on the scale of the dust, we expected the echo to be observed within a few months after the optical peak.
Assuming that most of the peak SN luminosity (with an energy of $\sim3\times10^{50}$ erg) is absorbed and reradiated during these months, we estimated that the echo has an IR luminosity of several $10^9$ $L_\odot$.
This hypothetical IR echo was not seen in the observations for PS15br on $+78$ day or SN 2017ens on $+169$ day.
Thus it is unlikely that such thick dust exists.
Therefore, we can rule out the shock emission heating model for PS15br and SN 2017ens.
While for SN 2017err, as no luminous IR echo was seen in the observation on $+147$ day, this model is also unlikely to hold.

\subsection{Newly Formed Dust?}
\label{sec:dust_newdust}

Using detailed multi-band data and model analysis, \cite{Chen2022} demonstrated that the IR excess of SN 2018bsz can be explained by a newly formed dust model.
SN 2017egm, SN 2017gci and SN 2018hti have similar LC and color evolution in the MIR band, so their IR excesses may also be explained by the same model.
Here we focus on the three SLSNe showing rebrightening, especially PS15br and SN 2017ens with good data quality.

A newly formed dust model can generally explain some MIR rebrightenings in normal SNe \citep[e.g. SN 2004dj,][]{Szalai2011}.
We noted that the energy released from the cooling of newly formed dust from the condensation temperature is far from sufficient to power the observed IR luminosity of PS15br and SN 2017ens on the order of $>10^8$ $L_\odot$ \citep[see a similar argument for SN 2005ip in][]{Fox2009}, and there must be an additional heating source.
As can be seen from Fig. 5, the dust luminosity is as high as $\sim10^9$ $L_\odot$ even at about $+600$ days, and drops no more than 0.3 dex between $+500$ and $+900$ days, which corresponds to a drop rate of $<0.2$ mag 100d$^{-1}$.
These can hardly be explained by heating by radiative decay of $^{56}$Co or a magnetar which spins down.
As we have seen features of strong CSM interaction in the two SLSNe, we inferred that the shock generated by the interaction can heat the dust.
Such a model is described in detail in \cite{Smith2008a} and \cite{Sarangi2018b}.
Here we briefly introduced the model with an illustration in Fig. 8.
The model assumes that dust forms in postshock clouds in a cool dense shell (CDS) located between the forward shock and the reverse shock.
After being compressed by the shock, the cloud's temperature drops from $\sim10^8$ K to $\sim10^4$ K within a few days as it radiates most of its initial internal energy.
The temperature then drops slowly because of additional heating by shock emission, including inward emission from the forward shock, and the outward emission from the reverse shock.
As the shock emission weakens and the CDS scales up in a few hundred days, the temperature in the core of the cloud drops to the condensation temperature, and dust begins to form.
The newly formed dust is heated by shock emission in two ways.
On one hand, the UV/optical emission is directly absorbed by the dust.
On the other hand, the far-UV and soft X-ray emission heats the ambient gas, and indirectly heats the dust via reradiation from the gas or collision by the gas.
The heated newly formed dust explains the observed IR emission at the late time.
To check whether such a model is suitable for the two SLSNe, we made the following four tests.

\begin{figure}
\centering{
  \includegraphics[scale=0.8]{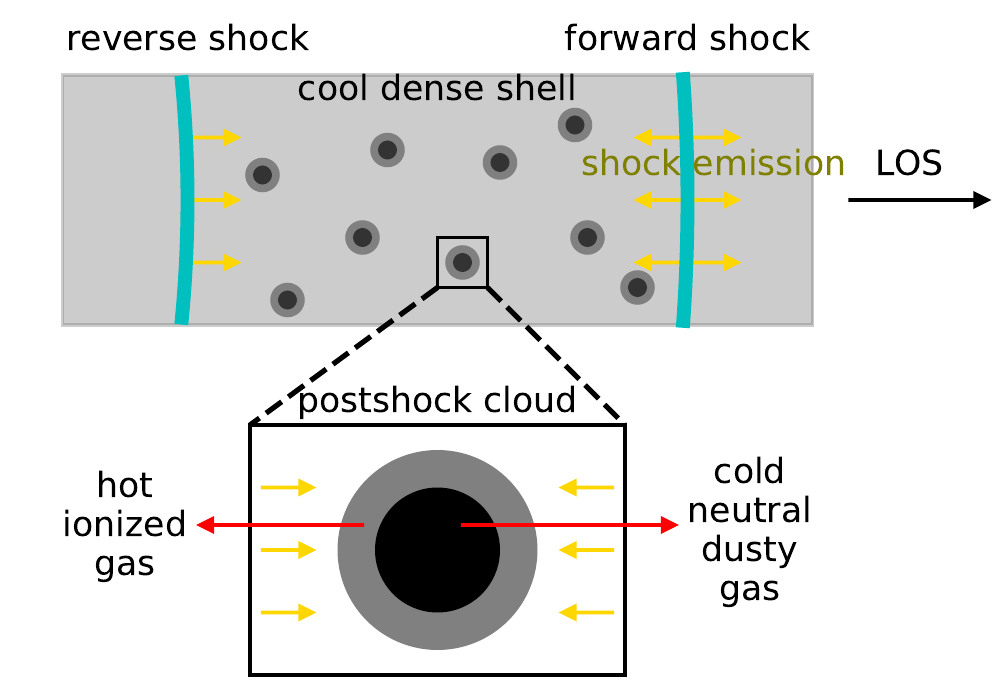}
  \caption{
  An illustration of the newly formed dust model for late-time MIR emission in some SNe with strong CSM interaction.}}
  \label{fig:illu_newdust}
\end{figure}

The first test is that the dust is located in a spherical shell behind the forward shock.
That is, $r_{\rm dust}<r_{\rm FS}$, where $r_{\rm dust}$ is the outer radius of this shell and $r_{\rm FS}$ is the radius of the forward shock.
We estimated the $r_{\rm dust}$ in PS15br and SN 2017ens using blackbody radius $r_{\rm bb}$ obtained by fitting the SED with blackbody curves, which is shown in Fig. 9(a).
As can be seen in Fig. 9(a), $r_{\rm bb}<r_{\rm FS}$ requires a shock velocity of at least $\sim10000$ km s$^{-1}$ assuming homogeneous expansions.
Such values are possible as SLSNe typically show higher shock velocities than normal SNe \citep{Inserra2018,Konyves-Toth2021}.

The second test is about energy balance: the observed dust luminosity cannot exceed the heating power at a certain time.
In Fig. 5, we compare the observed $L_{\rm dust}$ curves and the predicted radiant power of the shock ($L_{\rm shock}$) for PS15br and SN 2017ens (see Appendix E for the details of the model, note that $L_{\rm bol}$ is dominated by $L_{\rm shock}$ after 200 days).
We found that after around $+$500d, the MIR LC declines in step with $L_{\rm shock}$.
This is evidence that the dust luminosity is limited by the heating power.
Assuming amorphous carbon or silicate dust, the dust luminosity is roughly consistent with the predicted $L_{\rm shock}$ considering the uncertainties in the models.

The third test is that the postshock clouds can be cool enough for the dust to form.
Using abundant multi-band data, \cite{Sarangi2018b} estimated how $L_{\rm shock}$ and the scale of CDS vary with time in SN 2010jl, and inferred that dust can form 380 days after the explosion.
We assumed that the dust can also form in the two SLSNe when the similar $L_{\rm shock}/r_{\rm FS}^2$ values are obtained.
The estimated $L_{\rm shock}$ of $\sim10^9$ $L_\odot$ in PS15br and SN 2017ens at around $+$400 days is similar to that in SN 2010jl at similar epochs.
If assuming $r_{\rm FS}$ close to that of SN 2010jl at similar epochs, the dust can form in the two SLSNe at around 400 days.
Moreover, if assuming a larger $r_{\rm FS}$, as suggested by larger $r_{\rm BB}$ at around $+$600 days ($\sim6\times10^{16}$ cm for PS15br and SN 2017ens and $\sim3\times10^{16}$ cm for SN 2010jl), the dust can form earlier.
The predicted onset of dust formation, $+$300 $\sim$ $+$400 days, is consistent with the observed time of MIR rebrightening.

The final test is about the required dust mass.
In Fig. 9(b) we show the dust mass $M_d$ in PS15br and SN 2017ens, which was obtained in Section 3.2.2 from SED fitting by assuming optically thin dust with grain radius of 0.1 $\mu$m.
One should treat these values with caution because measurements of $M_d$ in SNe are usually model-dependent.
We examined the uncertainty that may arise from different model assumptions.
The optical depth at $\sim$4 $\mu$m of a homogeneous dust sphere with a radius $r$ can be expressed as:
\begin{equation}
\tau(\lambda) = \frac{ 3\kappa(\lambda) M_d }{ 4\pi r^2 } = \tau_0 \left( \frac{M_d}{10^{-2}\ M_\odot} \right) \left( \frac{r}{10^{17}\ \rm cm} \right)^{-2}
\end{equation}
where $\tau_0$ is 0.4 and 0.13 for amorphous carbon and silicate dust, respectively.
Using the measured $M_d$ for PS15br and SN 2017ens and assuming $r = r_{\rm bb} \sim 10^{16.8}$ cm, the optical depth is about 1, and hence the optically thin assumption that we used may hold.
However, optically thick dust is possible.
If so, the $M_d$ estimated earlier is only the mass in the $\tau<1$ layer, and the total $M_d$ needs to be multiplied by the IR optical depth.
In addition, if different grain radii are assumed, the resulting $M_d$ is also different.
Following some previous works \citep[e.g.,][]{Stritzinger2012}, we re-measured the $M_d$ assuming grain radii of 0.01 and 1 $\mu$m, and found several times differences.
In summary, our $M_d$ measurements can be conservative order-of-magnitude estimations.
We also show the measurements of $M_d$ for SN 1987A \citep{Wesson2015}, SN 2010jl \citep{Gall2014} and SN 2018bsz \citep{Chen2022} in Fig. 9(b) for a comparison.
The $M_d$ in PS15br and SN 2017ens at around $+$600d are about one order of magnitude higher than the values of SN 1987A and SN 2010jl at similar epochs, and nevertheless is within the scope of dust formation theory \citep[e.g.,][]{Sarangi2018a}.
In particular, the $M_d$ are consistent with that of SN 2018bsz in similar phases, which is predicted by the evolutionary trend assuming that $M_d\propto t^{2.4}$ \citep{Chen2022}.

\begin{figure}
\centering{
  \includegraphics[scale=0.8]{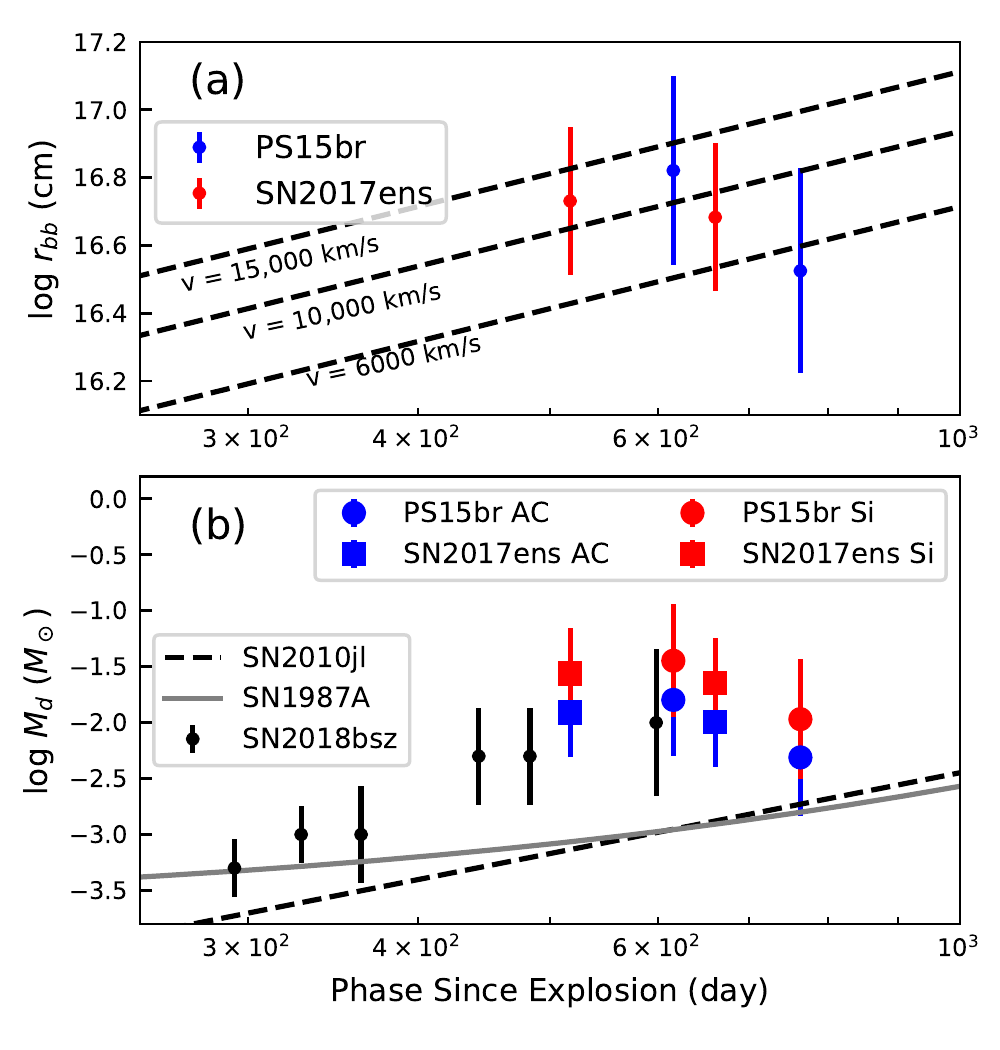}
  \caption{
  {\bf (a)}: The $r_{\rm bb}$ during the rebrightening in PS15br and SN 2017ens, and the shock expanding model that fit the data.
  In this plot, the phases are all relative to the explosion date.
  {\bf (b)}: The dust masses inferred for PS15br and SN 2017ens.
  We show the evolutions of $M_d$ of SN 1987A, SN 2010jl, and SN 2018bsz for a comparison.}}
  \label{fig:newdust}
\end{figure}

The newly formed dust model passes the tests described above.
Therefore we conclude that the late-time MIR emission of PS15br and SN2017ens can be explained by newly formed dust.
For SN2017err, although there is not enough data for the above tests, the late-time MIR emission may also be explained by the same model, because its MIR LC is similar to those of PS15br and SN 2017ens.

\subsection{A Summary on Dust Origin and Heating Mechanism}
\label{sec:dust_summary}

In SN 2017egm, SN 2017gci, SN 2018bsz and SN 2018hti, dust emission appears between $+$100d and $+$500d with a monochromatic luminosity of $10^7\sim10^8$ $L_\odot$.
In terms of time delay and luminosity, the dust emission can be explained by both the IR echo model and the newly formed dust model.
Distinguishing the two models may rely on multi-band monitoring data and detailed model analysis.
An example was presented in \cite{Chen2022} for SN2018bsz.
Chen et al. demonstrated that an IR echo model can hardly explain the dust emission.
The strongest evidence is that the optical monitoring did not detect scattered optical emission in step with the dust IR emission, which is predicted by the echo model.
Similar analysis can be made for other SLSNe to figure out the dust origin in the future.

In PS15br and SN2017ens, dust emission appears up to around $+$1000d with a luminosity of $10^8\sim10^9$ $L_\odot$.
The late-time emission can hardly be explained as an IR echo of the peak luminosity or an echo of the late-time shock emission, while can be naturally explained as the emission from newly formed dust.
Based on SED fittings, we estimated that dust with a mass of $\sim10^{-2}$ $M_\odot$ has been formed around 600 days after the explosion.
For SN 2017err, we failed to distinguish between IR echo and newly formed dust models because of insufficient data.
Anyhow, assuming that its late-time MIR emission comes from newly formed dust, a unified picture of the three SLSNe showing rebrightening features can be drawn.

\section{Discussions}
\label{sec:discussion}

\subsection{The Cause of The MIR Rebrightening}
\label{sec:discussion1}

What causes the MIR rebrightening by several hundred days in 3 out of 11 SLSNe?
We first check if this is related to the spectral types of the SLSNe, which were classified according to the near-peak spectra.
The three SLSNe, PS15br, SN 2017ens, and SN 2017err belong to type II, type I, and type IIn, respectively.
This implies that SLSNe of various spectral types may show MIR rebrightening.
The situation is similar to normal SNe: SNe with MIR rebrightening belongs to various spectral types such as Ia-CSM, Ib, IIP, and IIn \citep[e.g.,][]{Tinyanont2016,Szalai2019}.
So it is likely that the spectral type is not the determining factor for showing MIR rebrightening.

As we have pointed out in Section 4.3, all of the three SLSNe in the rebrightening group show features of strong interaction in their spectra.
A correlation between CSM interaction and late-time MIR rebrightening is also present in normal SNe.
For example, Ia-CSM and IIn are types of SNe that most frequently exhibit MIR rebrightening \citep[e.g.,][]{Fox2013,Szalai2019}.
Therefore it is likely that the MIR rebrightening in the three SLSNe is related to the CSM interaction.

The bolometric LCs of PS15br and SN 2017ens drop more slowly than other SLSNe in the late time \citep{Inserra2018,Chen2018}.
The best explanation for their slowly declining LCs is the radiant energy from shock, which is released during the CSM interaction.
Regardless of the energy source in the early time, the heating by interaction becomes dominant after about one year at the latest.
The emission from the reverse shock and the inward emission from the forward shock can be absorbed by the newly formed dust.
As the dust accumulates, an increasing proportion of the energy is transferred from the optical or X-ray bands to the IR band.
This explains the rise in the MIR LC.
When the amount of dust increases enough to convert most of the energy, the dust luminosity decreases slowly due to the limitation of the heating power.
This is consistent with the observation, as can be seen in Fig. 5.
This scenario can be tested by future observations with better data quality in the JWST era.

The situation is different for those SLSNe without strong interaction, such as SN2018bsz \citep{Chen2022}.
The newly formed dust, if present, can only be heated by radiative decay of $^{56}$Co or by magnetar.
The heating power drops off rapidly, and hence the emission from the newly formed dust is not strong enough to produce an observed rebrightening feature in the MIR LC.

\subsection{SLSNe as More Productive Dust Factories?}
\label{sec:discussion2}

\cite{Chen2022} demonstrated that SN 2018bsz possibly forms ten times more dust than normal CCSNe at similar epochs.
They further suggested that SLSNe may contribute $\sim$10\% of dust formation in the early universe with the following assumptions.
Firstly, all the SLSNe have high dust formation efficiencies.
Secondly, the final dust mass is also ten times different.
Thirdly, the proportion of SLSNe in CCSNe is 1\% in the early universe.
We found evidence that the dust mass in PS15br and SN2017ens is also ten times higher than normal CCSNe at the same epoch.
Our results appear to support the first hypothesis of \cite{Chen2022}.

However, more observations are required to conclude that SLSNe are more productive dust factories.
If the dust forms rapidly and is optically thick in a few years, as suggested by \cite{Dwek2019}, then the higher dust mass observed in SLSNe may simply be because the ejecta expands faster, causing more dust to be exposed.
Assuming that the newly formed dust in PS15br and SN2017ens is inside an expanding sphere, the required expansion velocity is as high as $\sim$10000 km s$^{-1}$, much higher than the typical value (2000--3000 km s$^{-1}$ for the inner ejecta) in normal CCSNe \citep[e.g.,][]{Szalai2013}.
A 3-fold difference in the ejecta velocity, and a corresponding 10-fold difference in the surface area of the dust sphere, is sufficient to explain the difference in the observed dust mass without additional assumptions about different dust formation efficiencies.
In addition, there is a possibility that the final dust mass is not greater in SLSNe though dust forms more rapidly in the first few years.
To check these possibilities and draw final conclusions on the dust formation efficiency in SLSNe require longer monitorings up to decades in the future.

\subsection{Is There an Upper Limit to The MIR Luminosity of SNe?}
\label{sec:discussion3}

\cite{Szalai2019} collated the Spitzer observations of 693 SNe with known redshift.
They found that the most MIR luminous SNe are of the IIn and Ia-CSM types, although these are only a small proportion of the total SNe.
Among the sample of Szalai et al., the SNe with the highest MIR luminosity is SN 2010jl and SN 2013cj, both of which are of the IIn type.
The absolute magnitudes of these two SNe are around $-22.2$ at 3.6 $\mu$m and around $-22.8$ at 4.5 $\mu$m.
\cite{Jiang2019} examined the WISE data of some known SNe with small angular separations from the galaxy centers.
They supplemented two normal SNe with luminosities similar to SN 2010jl, including SN 2013dz\footnote{Note that \cite{Jiang2019} misspelled it as SN2013dy} and SN 2014ab \citep[also reported in][]{Moriya2020}, which are also of the IIn type.
Interestingly, they found that ASASSN-15lh is extremely luminous in the MIR with W2-band absolute magnitude around $-25$, which is $>$2 magnitudes more luminous than the above-mentioned SNe IIn.
Note that the nature of ASASSN-15lh is unclear.
The discovery work \citep{Dong2016} considered it to be an SLSN, while some follow-up works \citep[e.g.,][]{Kruhler2018} considered it to be a TDE.

In the last decade, there have been a number of transient events occurring at the centers of galaxies that are difficult to categorize unambiguously.
These events include CSS100217:102913+404220 \citep{Drake2011,Zhang2022}, PS16dtm \citep{Blanchard2017,Jiang2017}, a flare in WISEA J094806.56+031801.7 \citep{Assef2018}, in addition to ASASSN-15lh.
All of these events have strong MIR emissions with peak W2-band absolute magnitudes between $-25$ and $-28$.
This makes an SNe identification controversial, as normal SNe is not so bright in the MIR band.
Moreover, IR flares with a week or no optical counterpart in the centers of some galaxies have recently been reported.
Examples are Arp-299B AT1 \citep{Mattila2018}, SDSS J165726.81+234528.1 \citep{Yang2019}, MCG-02-04-026 \citep{Sun2020}, AT 2017bgl \citep{Kool2020} and a sample of several tens MIR flares reported by \cite{Jiang2021}.
They were generally considered to be dust-obscured transient events, which may be TDE, SNe, or active galactic nuclei flares.
Although the nature of a few flares can be inferred from radio observations \citep[e.g.,][]{Mattila2018}, most flares have only IR information available.

The above situation leads one to an idea: is it possible to conclude that a flare is not an SN on the ground of too high MIR luminosity?
However, some previous works failed to fully address this issue due to the lack of a comprehensive understanding of the MIR emission of SLSNe.

The SLSNe in our sample have peak W2-band absolute magnitudes between $-21.5$ and $-23$.
They are not more MIR luminous than the most luminous SN IIn, though they are indeed more luminous than normal SNe on average.
Our results seem to support an upper limit on the MIR luminosity of SNe, though the sample size of 11 is not large.

A possible explanation is as follows.
MIR emission from SNe comes from IR echo or dust heated locally in the late time.
On one hand, the luminosity of the IR echo depends on the optical depth of the CSM dust to the SN UV/optical emission.
Echoing dust must be located outside the evaporation radius, which is expected to be between 0.1 and 1 light-year for the brightest SNe.
In addition, only echo from dust within a couple of light-years can emit significantly in the 3--5 $\mu$m band, and more distant dust produces an echo with lower luminosity or with SED peaks at longer wavelengths.
To make CSM dust thick enough to obscure the UV/optical emission, the required mass loss rate $\sim10^{-1} (f_d/0.01) (v_{\rm loss}/100\ {\rm km\ s^{-1}})$ $M_\odot$ yr$^{-1}$ can hardly be satisfied even by LBV stars.
Assuming that CSM dust absorbs 1\% to 10\% of the SNe UV/optical emission and that the duration of the IR echo is 10 times longer, a peak bolometric luminosity of $10^{11}$ $L_\odot$ only results in a dust luminosity of $10^8$--$10^9$ $L_\odot$.
On the other hand, the luminosity from locally heated dust is limited by both the size of the dust sphere and the heating power.
In the earlier time, $L_{\rm dust}$ does not exceeds $4\pi v_{\rm shock}^2 t^2 \sigma T_d^4$.
And when the dust sphere is large enough with the expansion of the shock, typically after a few hundred days, the heating power decreases because the central magnetar spins down or the density of interacting CSM decreases.
Therefore, there is an upper limit on the MIR luminosity of the emission produced by both mechanisms.


\section{Conclusions}
\label{sec:conclusion}

We have presented and analysed the MIR properties at 3.4 and 4.6 $\mu$m from a sample of 11 SLSNe, and investigated the dust emission in 7 of them.
The results are as follows.

1. MIR emission is detected from 10 out of 11 SLSNe with a threshold corresponding to luminosities at 3.4 and 4.6 $\mu$m of $\sim10^8$ $L_\odot$.
Most of the SLSNe show declining LCs and fade in about one year after the optical peak.
While the LCs of 3 SLSNe show a rebrightening in 1--2 years, and finally fade in 2--3 years.

2. Dust emission is detected in 7 SLSNe according to the MIR color of $W1-W2\sim1$ or the rebrightening feature in the LC.
For further study, we divided these 7 into two groups, a declining group including SN 2017egm, SN 2017gci, SN 2018bsz, and SN 2018hti, and a rebrightening group including PS15br, SN 2017ens, and SN 2017err.
For the declining group, the dust emission was detected about 100--500 days after the optical peak.
The dust emits in the MIR band with a monochromatic luminosity of $10^7\sim10^8$ $L_\odot$.
The dust masses may be $10^{-4}$--$10^{-2}$ $M_\odot$.
For the rebrightening group, the dust emission can be detected about 400--1000 days after the optical peak.
The dust luminosities are $10^8\sim10^9$ $L_\odot$ and the total energies of dust emission are $\sim10^{50}$ erg.
The dust has a temperature of 600--1100 K, and its mass is $\sim1\times10^{-2}$ or $\sim3\times10^{-2}$ $M_\odot$ if assuming carbonaceous or silicate dust, respectively, with grain radii of 0.1 $\mu$m.

3. For the declining group, the dust emission can be explained using both an IR echo model and a newly formed dust model.
While for PS15br and SN 2017ens, the late-time (after $\sim$1 years) MIR emission can only be explained by a newly formed dust model.
In this model, dust with a mass of $\gtrsim10^{-2}$ $M_\odot$ has formed $\sim$600 days after the explosion.

4. Considering that all the three SLSNe in the rebrightening group show features of strong CSM interaction, we conclude that newly formed dust heated by interaction explains the late-time dust emission with $10^8\sim10^9$ $L_\odot$ and the rebrightenings in the LCs.
The absence of strong late-time dust emission in other SLSNe may be due to the lack of interaction to provide enough heating power.

5. Taking into account our measurements of dust formations for PS15br and SN 2017ens and the measurement for SN 2018bsz by \cite{Chen2022}, it seems that SLSNe form one order-of-magnitude more dust than normal SNe in the same phase.
However, to draw conclusions about whether SLSNe have higher dust formation efficiency, decades of monitoring are necessary.

6. We found that the MIR luminosity of SLSNe is on average higher than normal SNe.
However, except for ASASSN-15lh whose SLSN nature is under debate, SLSNe are not more MIR luminous than SNe IIn.
Thus MIR flares more luminous than SLSNe and SNe IIn are more likely to be associated with TDE or active galactic nuclei flares rather than SNe.

\section*{Acknowledgements}

We thank the anonymous referee for helping us improve the work.
We thank T. Chen, C. Inserra, and S. Smartt for providing us SLSNe data.
L. Sun acknowledges the support from National Natural Science Foundation of China (NFSC, grant No. 12103002), and from Anhui Provincial Natural Science Fondation (grant No. 2108085QA43).
L. Xiao is thankful for the support from NFSC (grant No. 12103050), Advanced Talents Incubation Program of the Hebei University, and Midwest Universities Comprehensive Strength Promotion project.
This publication makes use of data products from the Wide-field Infrared Survey Explorer, which is a joint project of the University of California, Los Angeles, and the Jet Propulsion Laboratory/California Institute of Technology, funded by the National Aeronautics and Space Administration.

\section*{Data Availability}
Any new data used in this paper is available in the Appendix.

\begin{appendix}

\section{The Optical Data of LSQ15abl and SN 2017err}

\begin{figure}
\centering{
 \includegraphics[scale=0.8]{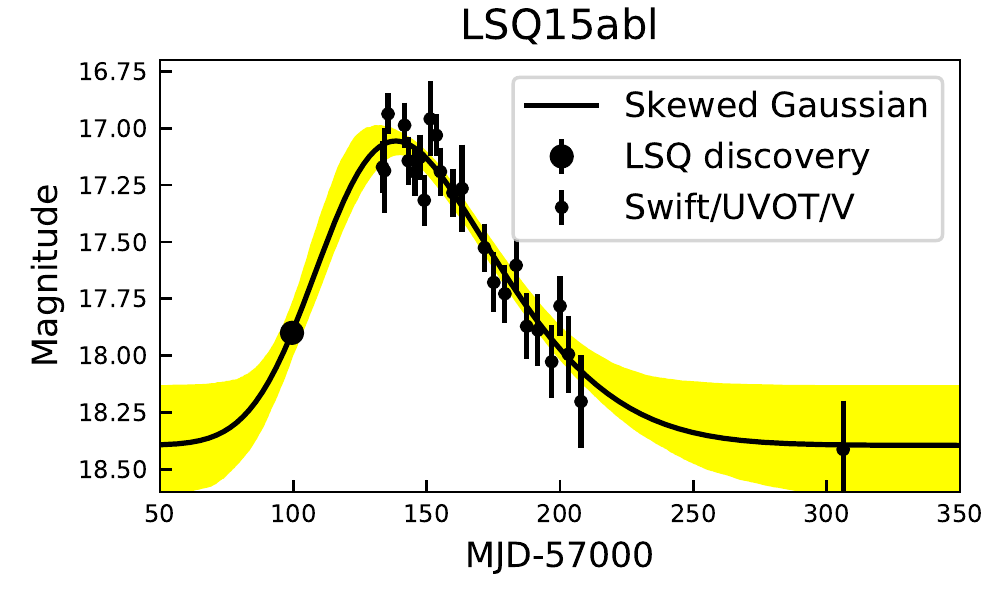}
 \caption{:
  The V-band LC of LSQ15abl and the skewed Gaussian model fitting to it.
  }}
\label{fig:lsq15abl}
\end{figure}

LSQ15abl was discovered on 2015 Mar 19 by La Silla-Quest survey \citep[LSQ,][]{Baltay2013}.
\cite{Prajs2015} identified it as SLSN-II, and measured a redshift of 0.087.
The SNe was monitored by Swift/UVOT between 2015 Apr 21 and 2015 Oct 10 in all six bands.
We show the V-band LC in Fig.~\ref{fig:lsq15abl}, using the LSQ magnitude reported in Prajs et al. and the Swift/UVOT magnitudes from Swift's Optical/Ultraviolet Supernova Archive \citep[SOUSA][]{Brown2014}.
The LC can be fitted using a skewed Gaussian curve, and with the curve, we measured a peak of the LC at MJD$=57136\pm5$ (2015 Apr 24).

SN 2017err was discovered on 2017 Jun 12 with o-band magnitude of $18.3\pm0.1$ \citep{Chen2017} by Asteroid Terrestrial-Impact Last Alert System \citep[ATLAS,][]{Tonry2011}.
From the ATLAS team (S. Smartt, private communication), SN2017err continued to rise to a measured brightest magnitude of $o=17.7\pm0.1$ on MJD$=57939$.
The ATLAS flux measurements and calibration are carried out on the difference images as described in \cite{Tonry2018} and \cite{Smith2020}.
Although given the faint magnitude of the host of $r=20.3$, there was little contamination at the peak.
After a gap in observing this position, ATLAS detected this object fading on MJD$=58054$ at $o=18.8\pm0.2$.
The ATLAS team reports that the gap in observing makes a true peak estimate uncertain, but it papers to be MJD$=57935\pm10$ days.
According to the optical/NIR follow-up observation taken on MJD$=57924$ ($g' = 17.92\pm0.07$, $r' = 18.01\pm0.10$, $i' = 18.07\pm0.07$, $z'=18.24\pm0.07$, $J = 18.96\pm0.13$, and $H = 19.14\pm0.19$, all in the AB system), we measured a blackbody temperature of SN 2017err of $12400\pm900$ K.
Using this temperature, the peak magnitude was converted to a bolometric luminosity of $1.1\times10^{44}$ erg s$^{-1}$.

\section{The Image Subtraction Results}

The reference, target, and difference images are shown in Fig.~\ref{figb1}.
The photometries are listed in Tab.~\ref{tabb1}.

\begin{table}
\scriptsize
\caption{WISE photometry of the 11 SLSNe.}
\begin{tabular}{cccccc}
\hline
\hline
MJD      & Phase   & \multicolumn{2}{c}{W1} & \multicolumn{2}{c}{W2} \\
         &         & flux     & mag    & flux     & mag \\
         & day     & $\mu$Jy  & Vega   & $\mu$Jy  & Vega \\
\hline
\multicolumn{6}{c}{PS15br} \\
\hline
57175.4 & +78.2   & 61(9) & 16.77(0.17) & 46(21) & $>$16.1 \\
57367.9 & +253.1  & 25(10) & $>$17.5 & 36(24) & $>$16.0 \\
57534.2 & +404.1  & 45(10) & 17.08(0.23) & 39(18) & $>$16.2 \\
57732.2 & +583.9  & 57(11) & 16.83(0.21) & 85(23) & 15.77(0.29) \\
57894.7 & +731.5  & 62(10) & 16.74(0.17) & 67(19) & 16.03(0.31) \\
58097.7 & +915.9  & 38(10) & 17.27(0.28) & 43(25) & $>$15.9 \\
58255.0 & +1058.8 & 20(11) & $>$17.4 & 30(21) & $>$16.1 \\
58463.4 & +1248.0 & 26(11) & $>$17.4 & 56(24) & $>$15.9 \\
58622.2 & +1392.3 & 10(12) & $>$17.4 & 13(23) & $>$16.0 \\
\hline
\multicolumn{6}{c}{SN2015bn} \\
\hline
57180.7 & +70.7   & 57(10) & 16.84(0.18) & 60(20) & 16.15(0.36) \\
57373.1 & +243.4  & 15(11) & $>$17.5 & 40(20) & $>$16.2 \\
57539.6 & +393.0  & 22(10) & $>$17.5 & 59(22) & $>$16.1 \\
57739.5 & +572.5  & 16(10) & $>$17.5 & 30(23) & $>$16.0 \\
57900.1 & +716.6  & 11(11) & $>$17.5 & 28(21) & $>$16.1 \\
\hline
\multicolumn{6}{c}{LSQ15abl} \\
\hline
57154.6 & +17.1   & 78(14) & 16.50(0.19) & 22(22) & $>$16.0 \\
57348.2 & +195.2  & 51(15) & 16.95(0.32) & 28(22) & $>$16.1 \\
57517.4 & +350.8  & 31(15) & $>$17.1 & 2(23) & $>$16.0 \\
57713.7 & +531.5  & 11(12) & $>$17.3 & 30(19) & $>$16.2 \\
57875.4 & +680.2  & -17(14) & $>$17.2 & 35(19) & $>$16.2 \\
\hline
\multicolumn{6}{c}{SN2016eay} \\
\hline
57524.5 & -15.4   & 27(9) & 17.64(0.36) & 23(18) & $>$16.2 \\
57724.4 & +166.2  & 34(11) & 17.41(0.34) & -2(25) & $>$15.9 \\
57886.4 & +313.3  & 11(10) & $>$17.5 & -5(23) & $>$16.0 \\
58092.0 & +499.9  & -1(11) & $>$17.4 & 38(23) & $>$16.0 \\
58246.9 & +640.6  & -10(10) & $>$17.5 & 13(23) & $>$16.0 \\
\hline
\multicolumn{6}{c}{SN2017ens} \\
\hline
57908.6 & -13.9   & 13(10) & $>$17.5 & 19(25) & $>$15.9 \\
58111.1 & +168.8  & 31(10) & 17.48(0.35) & 11(23) & $>$16.0 \\
58270.5 & +312.6  & 78(9) & 16.50(0.13) & 34(20) & $>$16.1 \\
58475.3 & +497.3  & 73(10) & 16.56(0.15) & 92(20) & 15.67(0.24) \\
58634.7 & +641.1  & 71(9) & 16.60(0.13) & 86(19) & 15.76(0.24) \\
58842.3 & +828.3  & 58(11) & 16.82(0.21) & 37(24) & $>$15.9 \\
59000.4 & +970.9  & 41(9) & 17.19(0.22) & 55(20) & $>$16.1 \\
\hline
\multicolumn{6}{c}{SN2017egm} \\
\hline
58067.5 & +137.5  & 81(19) & 16.46(0.26) & 166(30) & 15.04(0.20) \\
58227.3 & +292.5  & 81(19) & 16.45(0.26) & 105(32) & 15.54(0.33) \\
58433.1 & +492.2  & 69(23) & $>$16.6 & 110(31) & 15.48(0.30) \\
58589.3 & +643.8  & 12(20) & $>$16.8 & -4(33) & $>$15.6 \\
58797.4 & +845.7  & 14(19) & $>$16.8 & -4(32) & $>$15.6 \\
58954.9 & +998.5  & -11(20) & $>$16.8 & 26(34) & $>$15.6 \\
\hline
\multicolumn{6}{c}{SN2017err} \\
\hline
57894.9 & -36.2   & -4(9) & $>$17.7 & -6(19) & $>$16.2 \\
58097.9 & +147.2  & 16(8) & $>$17.8 & 19(23) & $>$16.0 \\
58255.3 & +289.3  & 40(8) & 17.22(0.22) & 32(19) & $>$16.2 \\
58463.6 & +477.5  & 44(12) & 17.11(0.28) & 66(23) & $>$16.0 \\
58622.4 & +620.9  & 30(10) & $>$17.5 & 51(20) & $>$16.2 \\
58827.9 & +806.6  & 10(11) & $>$17.5 & 7(22) & $>$16.0 \\
\hline
\multicolumn{6}{c}{SN2017gci} \\
\hline
58035.5 & +41.6   & 42(10) & 17.17(0.25) & 60(18) & 16.15(0.33) \\
58191.5 & +185.0  & 31(9) & 17.48(0.32) & 82(18) & 15.81(0.23) \\
58401.2 & +377.9  & 17(11) & $>$17.5 & 20(19) & $>$16.2 \\
58557.1 & +521.3  & -5(7) & $>$17.9 & -3(15) & $>$16.4 \\
58765.4 & +712.9  & 15(11) & $>$17.5 & -31(22) & $>$16.1 \\
\hline
\multicolumn{6}{c}{SN2018bgv} \\
\hline
58228.0 & -23.4   & 2(7) & $>$18.0 & -39(15) & $>$16.5 \\
58434.6 & +168.1  & 18(8) & $>$17.7 & 21(15) & $>$16.4 \\
58592.3 & +314.1  & -2(7) & $>$18.0 & -4(15) & $>$16.4 \\
58801.9 & +508.3  & 4(8) & $>$17.8 & -19(19) & $>$16.2 \\
58956.4 & +651.4  & 3(7) & $>$17.9 & 4(16) & $>$16.4 \\
\hline
\multicolumn{6}{c}{SN2018bsz} \\
\hline
58336.6 & +67.3   & 552(24) & 14.37(0.05) & 404(28) & 14.07(0.08) \\
58539.3 & +264.7  & 208(22) & 15.43(0.11) & 385(28) & 14.12(0.08) \\
58702.1 & +423.3  & 33(22) & $>$16.7 & 70(29) & $>$15.7 \\
58904.9 & +620.8  & 23(23) & $>$16.6 & -55(27) & $>$15.8 \\
59067.9 & +779.5  & -7(21) & $>$16.7 & -17(30) & $>$15.7 \\
\hline
\multicolumn{6}{c}{SN2018hti} \\
\hline
58508.2 & +41.1   & 300(11) & 15.03(0.04) & 177(21) & 14.97(0.13) \\
58715.5 & +236.4  & 75(13) & 16.54(0.18) & 75(26) & $>$15.9 \\
58872.4 & +384.3  & 39(10) & 17.25(0.27) & 68(19) & 16.01(0.31) \\
59080.7 & +580.6  & 21(10) & $>$17.6 & 29(22) & $>$16.0 \\
\hline
\end{tabular}
\label{tabb1}
\end{table}

\begin{figure*}
\centering{
 \includegraphics[scale=1.75]{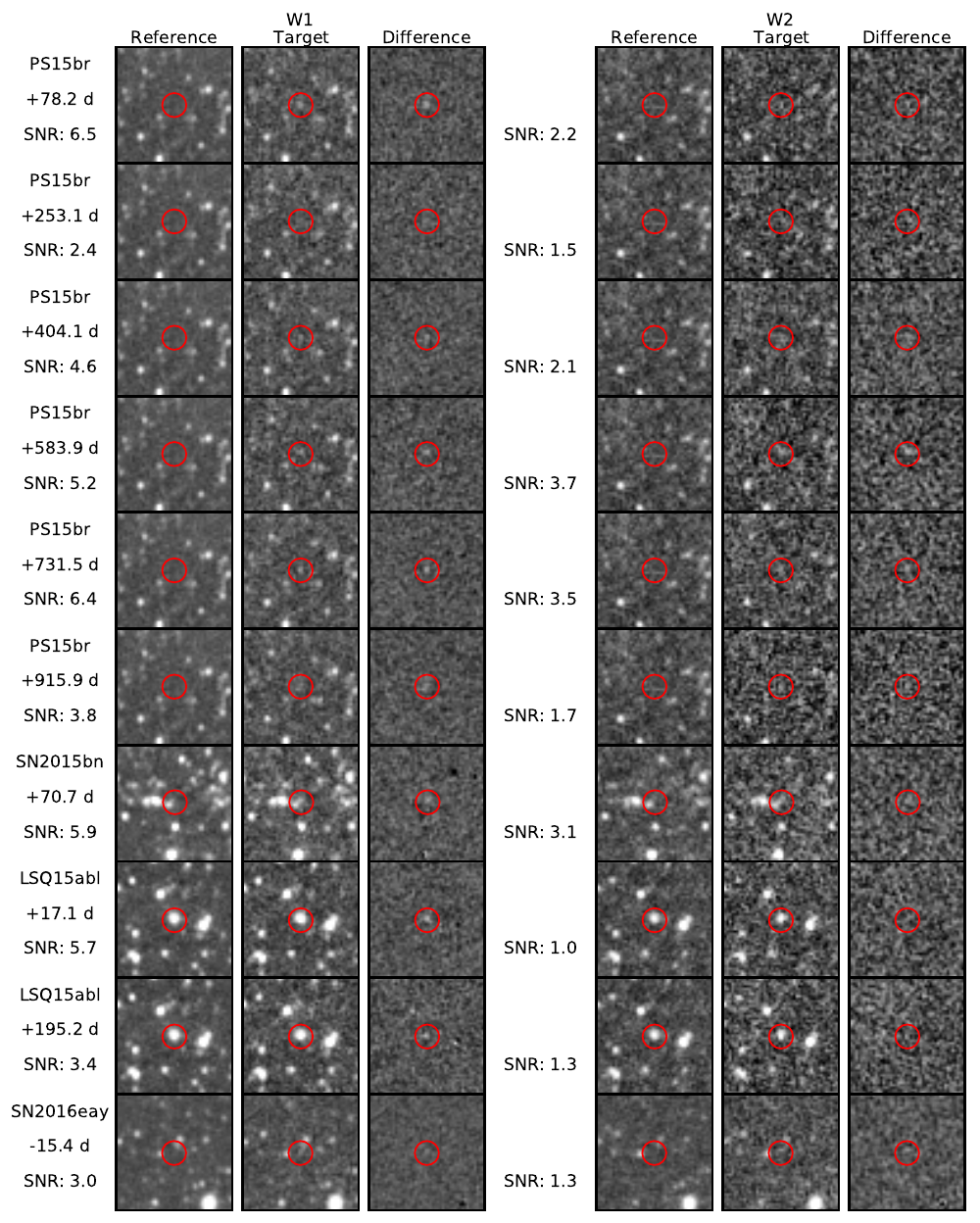}
 \caption{
  The reference, target, and difference images in the W1 band (left) and W2 band (right).
  Red circles show the target positions.
  The object name, the phase, and the signal to noise ratio are labeled.
  }}
\label{figb1}
\end{figure*}

\begin{figure*}
\centering{
 \includegraphics[scale=1.75]{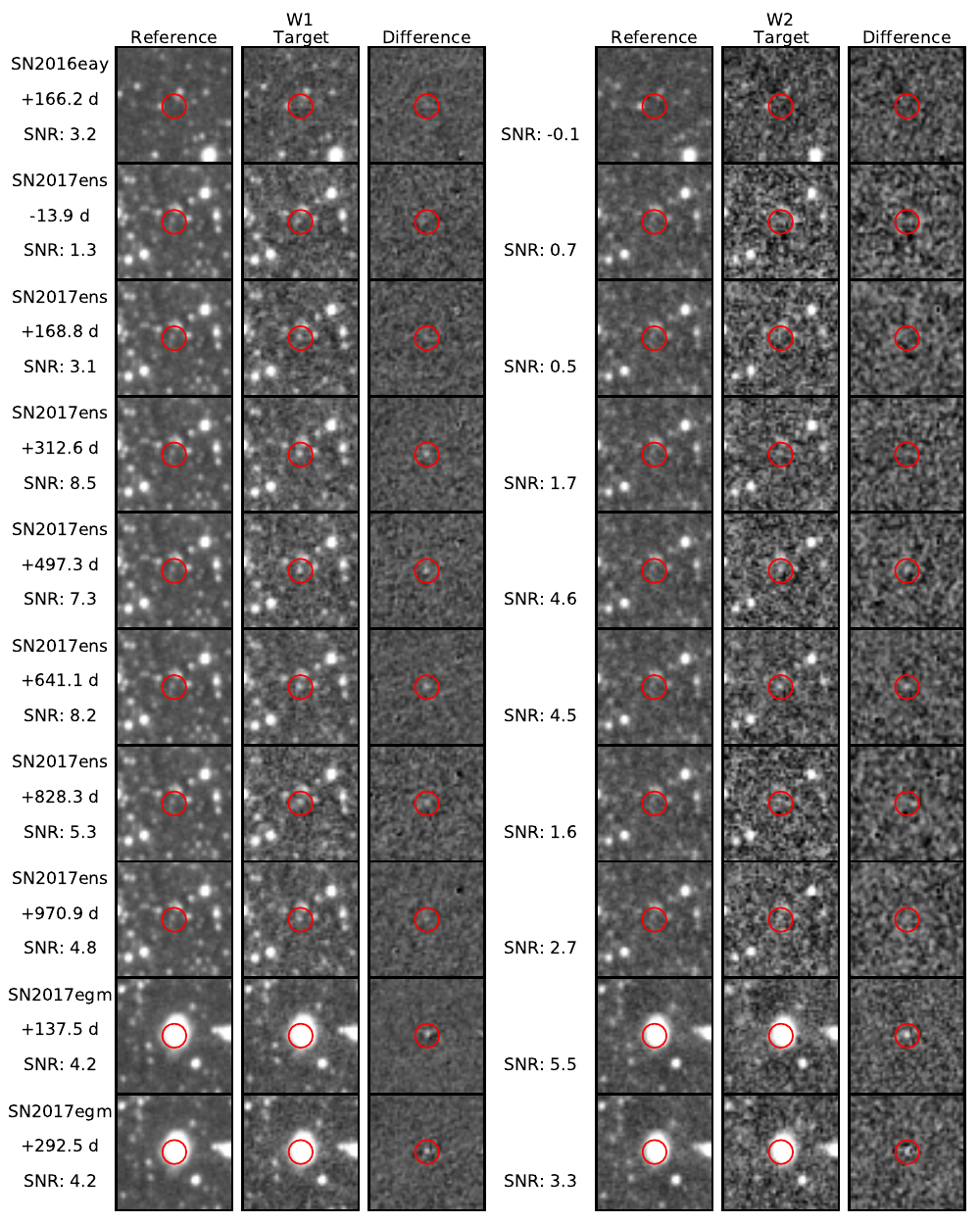}
 \caption{
  Continued to Fig.~\ref{figb1}.
  }}
\label{figb2}
\end{figure*}

\begin{figure*}
\centering{
 \includegraphics[scale=1.75]{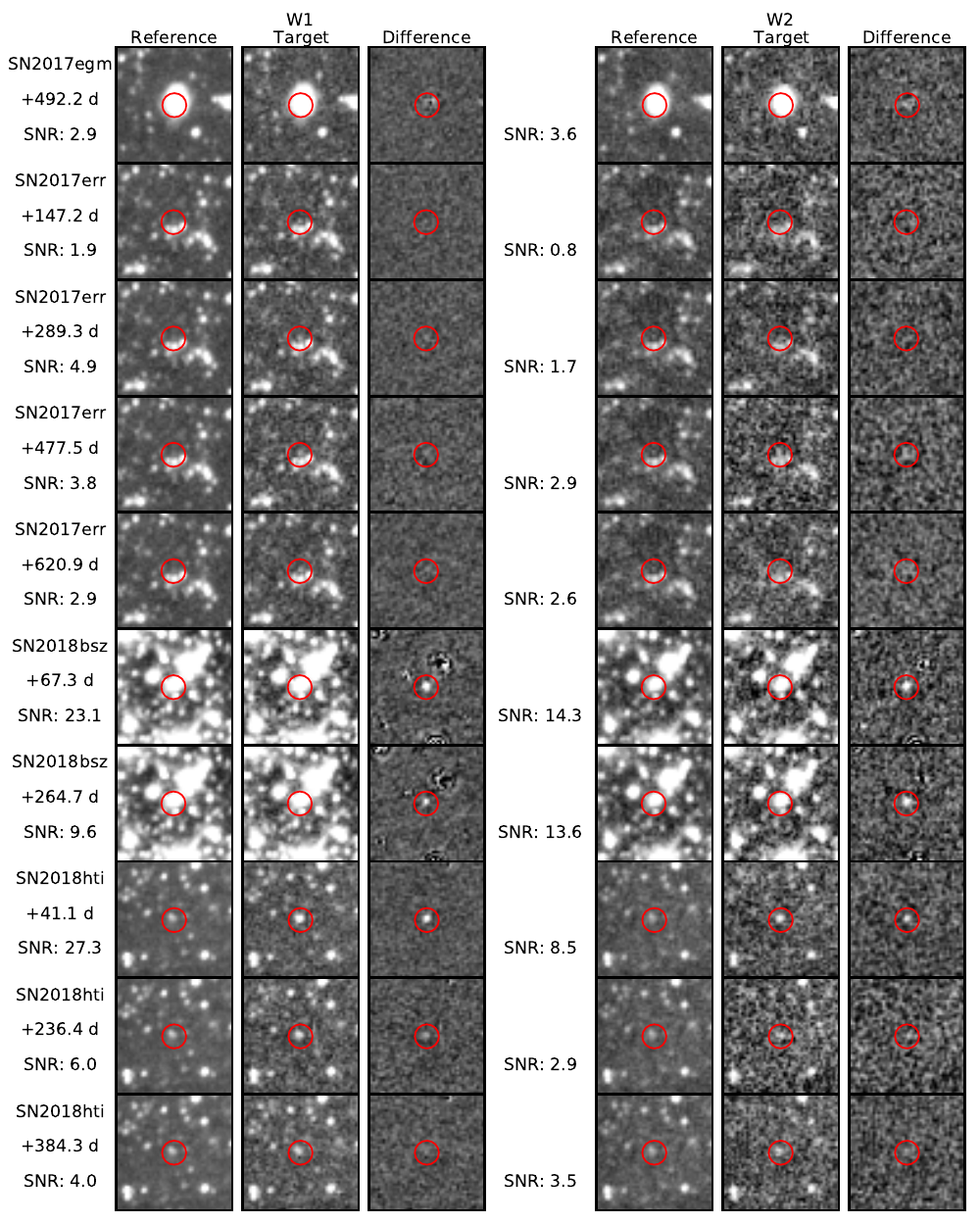}
 \caption{
  Continued to Fig.~\ref{figb1}.
  }}
\label{figb3}
\end{figure*}

\section{A Check of WISE Photometry of SN 2018bsz with Spitzer Result}

\begin{figure}
\centering{
 \includegraphics[scale=0.8]{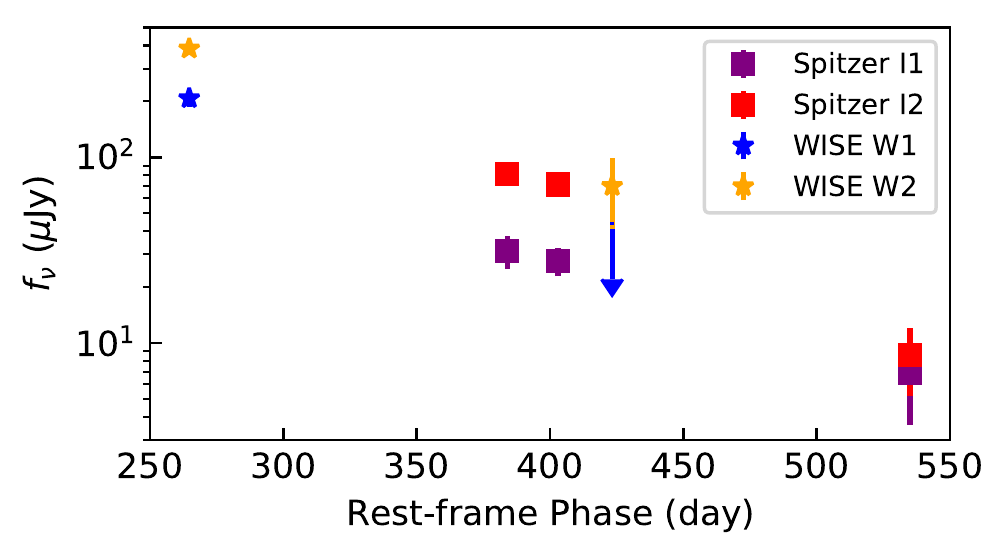}
 \caption{:
  A comparison between the WISE and Spitzer photometries of SN 2018bsz.
  }}
\label{figc1}
\end{figure}

SN 2018bsz is the only SLSNe in our sample which was observed by Spitzer.
IRAC observations were taken on $+384$, $+403$, $+535$ and $+564$ days with I1 and I2 filters.
Using images on $+564$ day as a reference, \cite{Chen2022} made difference images for the first three epochs, on which they measured the fluxes of SN 2018bsz.
We checked whether our measurements using WISE data are consistent with these measurements using Spitzer data.
As can be seen from Fig.~\ref{figc1}, the two sets of measurements agree with each other.

\section{Relation between WISE W1$-$W2 and Spitzer I1$-$I2 colors}

\begin{figure}
\centering{
 \includegraphics[scale=0.8]{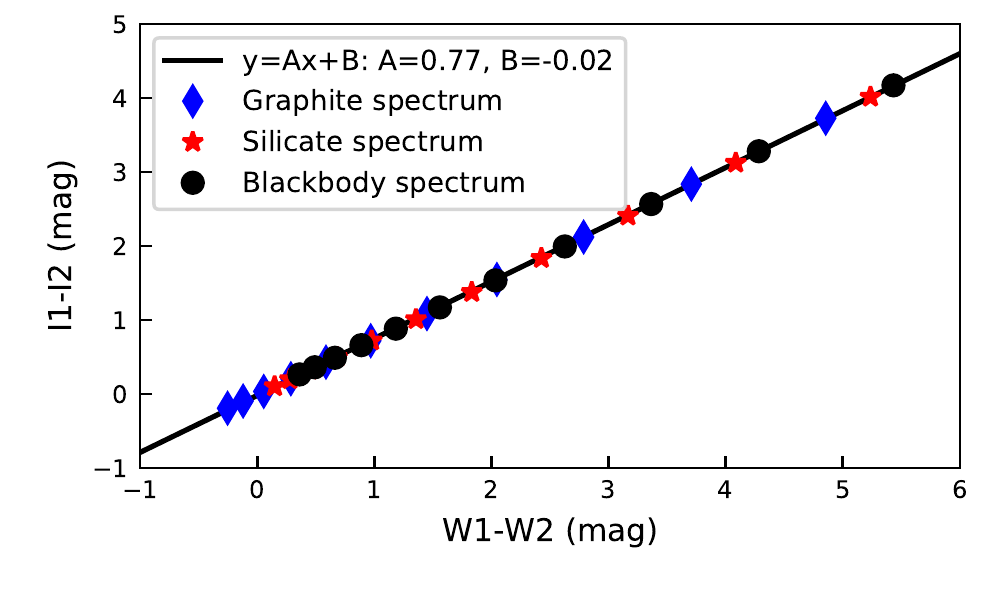}
 \caption{:
  The $W1-W2$ and $I1-I2$ colors for spectra of blackbodies and optically thin dust with temperatures between 200 and 2000 K, and the best-fitting linear function.
  }}
\label{figd1}
\end{figure}

In order to compare the SLSN $W1-W2$ color with the $I1-I2$ color of normal SNe measured by Spitzer in the literature, we search for the conversion between $W1-W2$ and $I1-I2$ colors for dust spectra.
We assumed a series of blackbody spectra with temperatures between 200 and 2000 K in 0.1 dex step, and calculated the synthetic magnitudes\footnote{We used the response curves of Spitzer filters from https://irsa.ipac.caltech.edu/data/SPITZER/docs/irac and those of WISE filters from https://wise2.ipac.caltech.edu/docs/release/prelim/expsup} in WISE and Spitzer bands, and hence the MIR colors.
We also calculated MIR colors for spectra of optically thin dust assuming graphite and silicate dust.
As can be seen in Fig.~\ref{figd1}, the $W1-W2$ color and the $I1-I2$ color are linear correlated for typical dust spectra.
By fitting these data points with a linear function, we obtained equation (1) as shown in the black line.

\section{The Bolometric Light Curves of PS15br and SN 2017ens}

\begin{figure}
\centering{
 \includegraphics[scale=0.8]{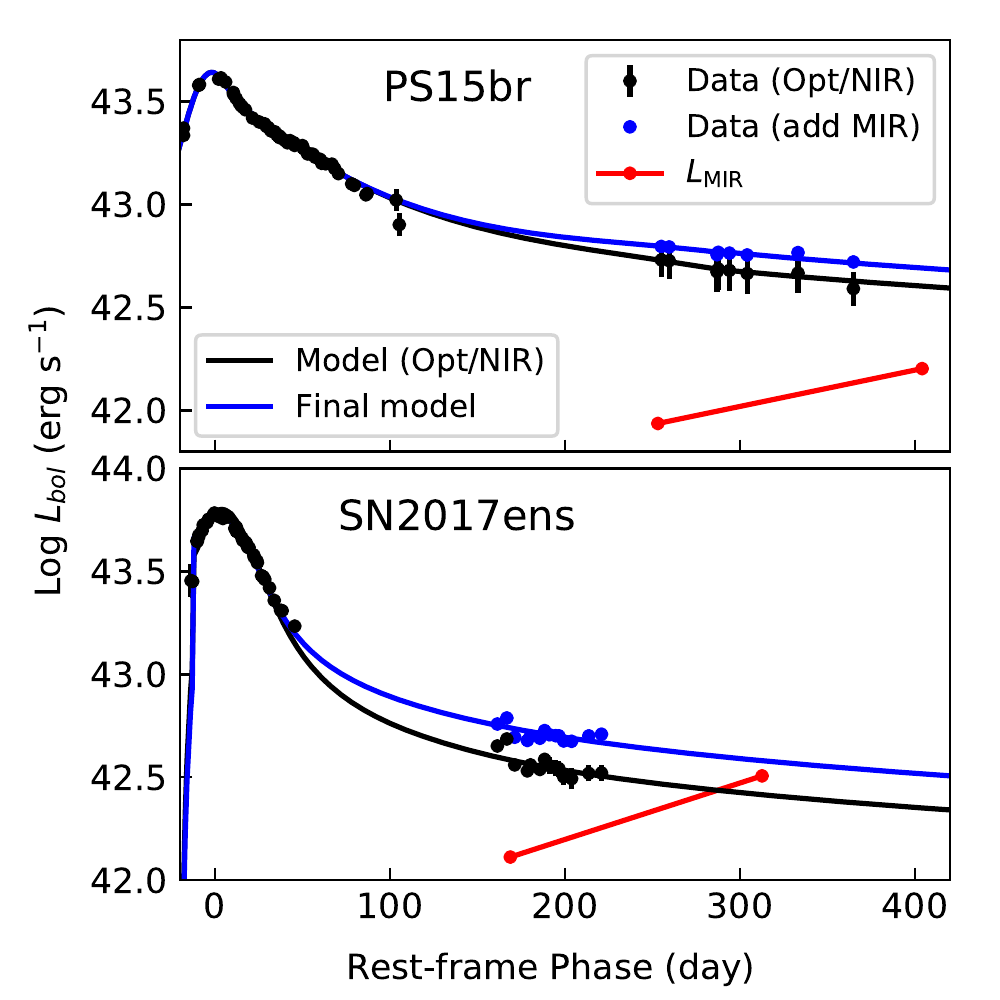}
 \caption{:
  The bolometric LC and models for PS15br and SN 2017ens.
  }}
\label{fige1}
\end{figure}

\cite{Inserra2018} and \cite{Chen2018} have calculated the bolometric LCs for PS15br and SN 2017ens, respectively.
However, they only considered UV, optical, and NIR emissions in their calculations.
By considering the MIR emission, we recalculated the bolometric LCs for PS15br and SN 2017ens following a method used in \cite{Chen2022}.
We collected the LC data from the literature, which are shown in Fig.~\ref{fige1} in black points.
The optical/NIR data are available until $+$365d for PS15br and until $+221$d for SN 2017ens.
We calculated MIR luminosities (red points) using the W1-band luminosity assuming a $C_B(W1)=1.5$.
We found that MIR emission accounted for more than 10\% of the total luminosity after $+$150d.
Therefore, for data points after that, we computed MIR luminosity by interpolating the observed data assuming an exponential function (red lines), and summed the optical/NIR luminosity and the MIR luminosity to obtain the bolometric luminosity (blue points).

The bolometric LCs of the two SLSNe decrease slowly in the late time.
This is likely because the heating of the CSM interaction dominates.
Therefore, we extrapolated the measured bolometric LC to later phases using a CSM interaction model.
Following \cite{Chen2018}, we assumed a function of $L_{\rm bol}=L_{\rm int,1} (t/{\rm 1\ s})^{-3/5}$ for this model.
In addition, we concatenated the early-time data simply using cubic spline functions.
The final results are shown in blue lines.

\section{The Bolometric Corrections}

\begin{figure}
\centering{
  \includegraphics[scale=0.8]{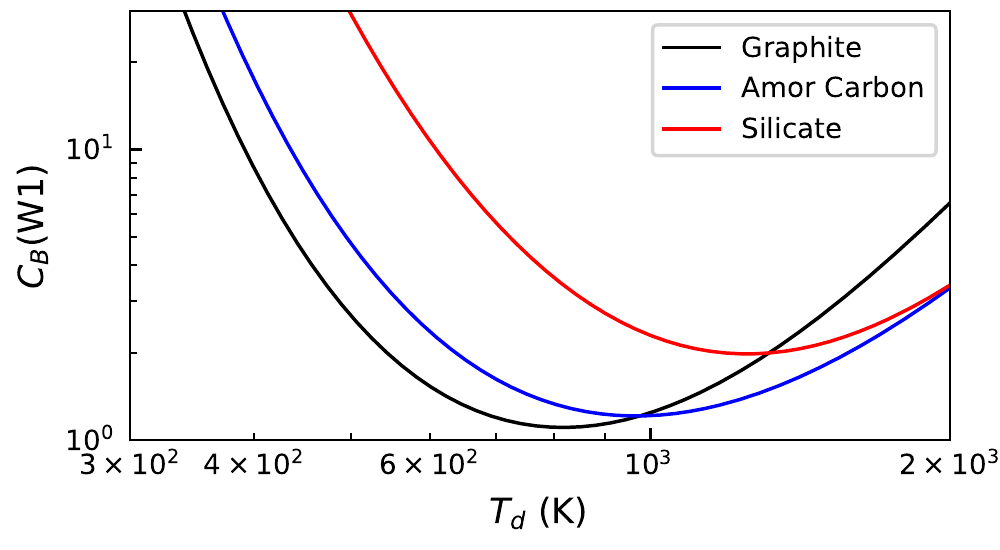}
  \caption{The bolometric correction in the W1 band as a function of $T_d$ at $z=0.105$.}}
  \label{figf1}
\end{figure}

We calculated the bolometric corrections $C_B$ in the W1-band for different dust temperatures $T_d$ using equation (3) by assuming dust spectra expressed as equation (2).
The dust spectra were shifted to the observer's frame according to $z=0.105$, since PS15br, SN 2017ens and SN 2017err are at redshifts between 0.101 and 0.109.
The results, assuming three different dust compositions, are displayed in Fig.~\ref{figf1}.
The minimum values of $C_B$ and the corresponding $T_d$ are listed in Tab.~\ref{tabf1}.

\begin{table}
\footnotesize
\caption{Minimum $C_B$ in the W1 band at $z=0.105$}
\begin{tabular}{cccccccc}
\hline
Dust        & minimum $C_B$\\
\hline
Graphite    & 1.10 (810 K)\\
Amor Carbon & 1.21 (970 K)\\
Silicate    & 1.98 (1250 K)\\
\hline
\end{tabular}
\label{tabf1}
\end{table}

\end{appendix}

\bsp	
\label{lastpage}
\end{document}